\documentclass{emulateapj}

\newcommand{\simgt}{{\raise-.5ex\hbox{$\buildrel>\over\sim$}}}
\newcommand{\simlt}{{\raise-.5ex\hbox{$\buildrel<\over\sim$}}}
\newenvironment{packed_enum}{
\begin{enumerate}
  \setlength{\itemsep}{1pt}
  \setlength{\parskip}{0pt}
  \setlength{\parsep}{0pt}
}{\end{enumerate}}

\slugcomment{}

\shorttitle{II. Color Trends and Mass Profiles}
\shortauthors{Herrmann, Hunter, \& Elmegreen}

\begin{document}

\title{Surface Brightness Profiles of Dwarf Galaxies: II. Color Trends and Mass Profiles}

\author{Kimberly A.\,Herrmann$^1$, Deidre A.\,Hunter$^2$, \& Bruce G.\,Elmegreen$^3$}
\affil{$^1$Penn State Mont Alto, 1 Campus Drive, Mont Alto, PA 17237, USA; kah259@psu.edu}
\affil{$^2$Lowell Observatory, 1400 West Mars Hill Road, Flagstaff, AZ  86001, USA; dah@lowell.edu}
\affil{$^3$IBM T.\,J.\,Watson Research Center, 1101 Kitchawan Road, Yorktown Hts., NY  10598, USA; bge@us.ibm.com}

\begin{abstract}
In this second paper of a series, we explore the $B-V$, $U-B$, and FUV$-$NUV radial color trends from a multi-wavelength sample of 141 dwarf disk galaxies.  Like spirals, dwarf galaxies have three types of radial surface brightness profiles: (I) single exponential throughout the observed extent (the minority), (II) down-bending (the majority), and (III) up-bending.  We find that colors of (1) Type~I dwarfs generally become redder with increasing radius unlike spirals that have a blueing trend that flattens beyond $\sim$1.5 disk scale lengths, (2) Type~II dwarfs come in six different ``flavors,'' one of which mimics the ``U'' shape of spirals, and (3) Type~III dwarfs have a stretched ``SÕ'' shape where central colors are flattish, become steeply redder to the surface brightness break, then remain roughly constant beyond, similar to spiral Type~III color profiles, but without the central outward bluing.  Faint ($-9 > M_B > -14$) Type~II dwarfs tend to have continuously red or ``U'' shaped colors and steeper color slopes than bright ($-14 > M_B > -19$) Type~II dwarfs, which additionally have colors that become bluer or remain constant with increasing radius.  Sm dwarfs and BCDs tend to have at least some blue and red radial color trend, respectively.  Additionally, we determine stellar surface mass density ($\Sigma$) profiles and use them to show that the break in $\Sigma$ generally {\it remains} in Type~II dwarfs (unlike Type~II spirals) but generally {\it disappears} in Type~III dwarfs (unlike Type~III spirals).  Moreover, the break in $\Sigma$ is strong, intermediate, and weak in faint dwarfs, bright dwarfs, and spirals, respectively, indicating that $\Sigma$ may straighten with increasing galaxy mass.  Lastly, the average stellar surface mass density at the surface brightness break is roughly 1$-$2~$M_{\odot}$~pc$^{-2}$ for Type~II dwarfs but higher at 5.9~$M_{\odot}$~pc$^{-2}$ or 27~$M_{\odot}$~pc$^{-2}$ for Type~III BCDs and dIms, respectively.
\end{abstract}


\section{INTRODUCTION}
To first order, disk light in galaxies falls off exponentially with increasing radius; why this is so is still unknown.  However, large and deep photometric studies (Erwin et al.\,2005; Pohlen \& Trujillo 2006; Erwin et al.\,2008; Guti\'{e}rrez et al.\,2011) have revealed {\it breaks} in the light fall off in many spiral galaxies such that the light beyond the break falls off more steeply (Type~II, down-bending; Pohlen \& Trujillo 2006) or less steeply (Type~III, up-bending; Erwin et al.\,2005).  In late-type spirals, galaxies with light falling off with a single exponential throughout the whole probed disk (Type~I profiles) are the minority whereas Type~II profiles are the majority (Pohlen \& Trujillo 2006).

Bakos et al.\,(2008; hereafter BTP08) examined radial surface brightness and $g'-r'$ color profiles from 69 spiral galaxies with deep SDSS $g'$ and $r'$ photometry and found different color trends characteristic of each type, especially after the radii were scaled by 2.5 times the scale length for Type~I profiles and the break scale length for Type~II and III profiles.  We digitized the averaged results from Figure~1 of BTP08 and present them in gold in Figures~1 and 2.  In spirals, all three profile types have a blue trend from the center outward typically attributed to a radial decrease in age and metallicity (de Jong 1996) but (1) Type~I colors then stay roughly constant beyond 1.5 scale lengths, (2) Type~II colors get redder beyond the break (having an overall ``U'' shape), and (3) Type~III profiles reach their bluest colors at roughly half the break radius, then become redder to the break, and then stay roughly constant beyond.  Without the inner blue trend, the color profiles for Type~III spirals are roughly ``S''-shaped, especially an ``S'' stretched horizontally with the top to the right and the bottom to the left so it does not curve back on itself.  When the ``U''-shaped color trend of Type~II spirals was used to estimate a changing mass-to-light ratio ($M/L$), the break in the resulting surface mass density profile was remarkably reduced (BTP08).  Their interpretation for Type~II spirals is that the color change is caused by differing stellar populations from older, redder stars migrating outward due to interactions with spiral arms.  Some simulations support this interpretation \citep{r+08a,r+08b,ms+09}, but others indicate that a disk-metallicity gradient and an inside-out formation scenario might also be important \citep{sb+09}.

\begin{figure*}
\epsscale{1.05}
\plotone{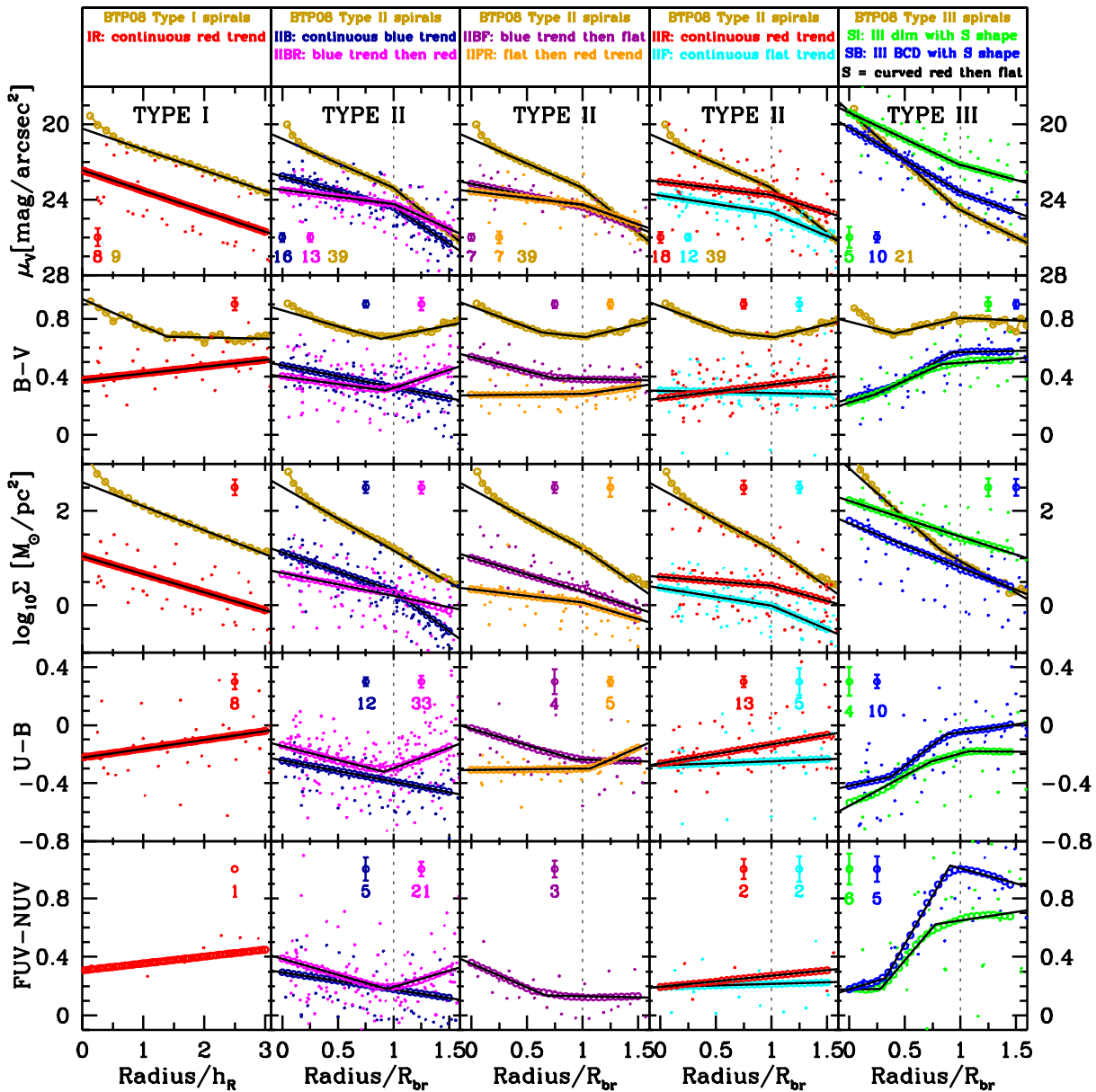}
\caption{Scaled and averaged profiles for the {\it broken-by-default} sample (broken-by-default means that galaxies which had profiles that could be interpreted as either one component or two components were given two components in this sample): {\it Top row}: $V$-band radial surface brightness profiles; {\it $2^{nd}$ row}: $B-V$ radial color profiles; {\it $3^{rd}$ row}: Surface mass density profiles obtained from applying a color-to-$M/L$ relationship (Herrmann et al., in preparation); {\it $4^{th}$ row}: $U-B$ radial color profiles; {\it Bottom row}: FUV$-$NUV radial color profiles.  The error bars show the average uncertainty in the mean, i.e., $\sigma/\sqrt{N}$ where $\sigma$ is the standard deviation and $N$ (shown below each error bar) is the number of galaxies used to calculate the average.  The small dots are the individual galaxy data points.  The gold data points show spiral results digitized from BTP08.  The parameters for the linear (or broken linear) fits indicated by the black lines are given in Table~3.  Note that the dwarf Type~II sample has been split between three columns for clarity such that the 2nd column contains IIB and IIBR dwarfs, the 3rd column contains IIBF and IIFR dwarfs, and the 4th column contains IIR and IIF dwarfs. \label{Brk_ave} }
\end{figure*}

\begin{figure*}
\epsscale{0.88}
\plotone{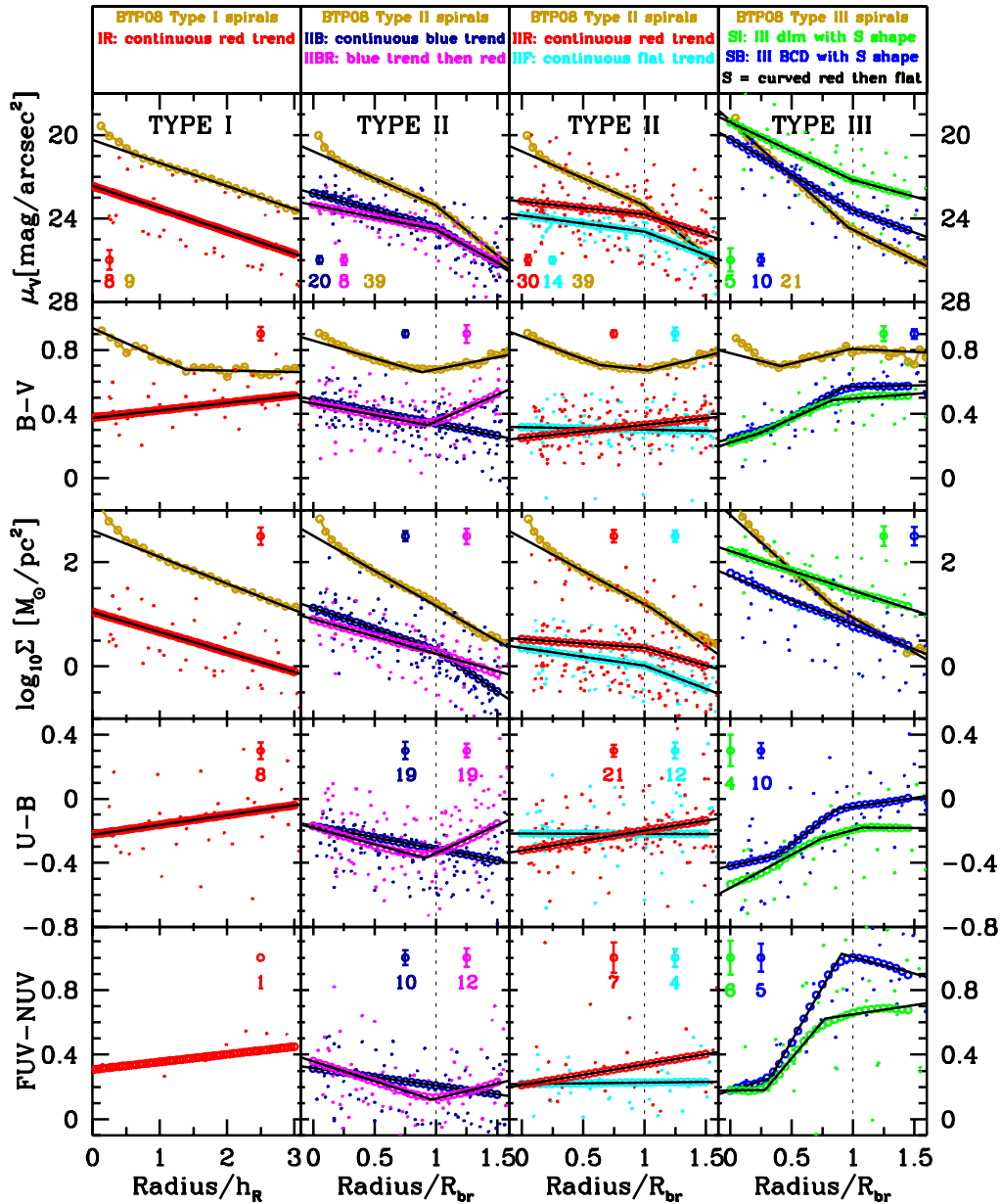}
\caption{Scaled and averaged profiles for the {\it single-by-default} sample.  Since some of the color profiles were well fit by both a single line and a broken (two line) fit, the analysis was performed twice.  All clear classifications were either single or broken in both analyses, but the ambiguous profiles were fit with a single line (``single-by-default'') or a broken fit (``broken-by-default'') in the two separate analyses.  Note that: (1) the Type~I and III columns as well as the gold spiral curves in the 2nd and 3rd columns are exactly the same as in Figure~1 and (2) virtually all the IIBF (blue trend then flat) and IIFR (flat then red trend) classifications were ambiguous, so IIBF and IIFR color flavors have disappeared in the single-by-default analysis results.  See the caption of Figure~1 for more information about the plot.\label{Sgl_ave} }
\end{figure*}

\begin{deluxetable*}{cccccccccccc}
\tabletypesize{\scriptsize}
\tablecaption{Galaxy Sample and Color Profile Classifications \label{Sample}}
\tablewidth{0pt}
\tablehead{\colhead{Galaxy} & \colhead{Group\tablenotemark{a}} & \colhead{$M_B$\tablenotemark{b}} & \colhead{$b/a$\tablenotemark{c}} & \colhead{Type\tablenotemark{d}} & \colhead{$B-V$\tablenotemark{e}} &\colhead{O/S\tablenotemark{f}} & \colhead{$U-B$\tablenotemark{e}} & \colhead{O/S\tablenotemark{f}} & \colhead{FUV-NUV\tablenotemark{e}} & \colhead{O/S\tablenotemark{f}} & \colhead{Notes\tablenotemark{g}}}
\startdata
0467-074 & Im & -18.31 & 0.85 & II & IIBF/B & \dots & IIF & \dots & \dots & \dots & \dots \\
1397-049 & Im & -17.23 & 0.74 & II & IIB & \dots & \dots & \dots & \dots & \dots & \dots \\
A1004+10 & Im & -15.60 & 0.63 & III & SI & S & \dots & \dots & \dots & \dots & \dots \\
A2228+33 & Im & -17.24 & 0.92 & III+II & IIFR/F & S & IIF & S & \dots & \dots & B \\
CVnIdwA	 & Im & -12.16 & 0.78 & FI & IIBR & \dots & IIBR & \dots & IIBR & \dots & LT
\enddata
\tablenotetext{}{Note: Table~1 is published in its entirety in the electronic edition of the {\it Astronomical Journal}.  A portion is shown here for guidance regarding its form and content.}
\tablenotetext{a}{Morphological Hubble types are from de Vaucouleurs et al.\,(1991; RC3).  If no type is given there, we used types from NED.}
\tablenotetext{b}{Absolute magnitude in $B$-band, calculated from Hunter \& Elmegreen (2006) using distances from Paper~I.  ``NB" = no $B$ data.}
\tablenotetext{c}{Minor-to-major axis ratios, $b/a$, from Hunter \& Elmegreen (2006) used for inclination corrections assuming $b/a_0 = 0.3$.}
\tablenotetext{d}{Surface brightness profile type per galaxy from Paper~I.  Galaxies with two breaks are listed as ``inner type + outer type'' and ones with different types depending on wavelength are generally listed blue to red separated by a ``/".}
\tablenotetext{e}{Color profile classification from this study.  The presence of a ``/'' indicates an ambiguous profile; the first classification was used in the broken-by-default analysis and the second in the single-by-default analysis.}
\tablenotetext{f}{Outliers (``O'') and profiles that were too short (``S''); both were not used in the averaging analyses.}
\tablenotetext{g}{Notes column: B = barred, PM = peculiar morphology, LT = LITTLE THINGS (Hunter et~al.\,2012) member, T = THINGS \citep{THINGS} member, VO = $V$-band only}
\end{deluxetable*}

Local spirals are not the only galaxies known to have Type~I, II, and III profiles: galaxies at intermediate redshifts ($0.1 < z < 1.1$; P\'{e}rez 2004; Azzollini et~al.\,2008; Trujillo et~al.\,2009) and local dwarf galaxies (Hunter \& Elmegreen 2006; Zhang et~al.\,2012a; Herrmann et~al.\,2013) also have single and broken surface brightness profiles.  Additionally, dwarfs have a special case of Type~II profiles unobserved in spirals: FI, flat or even increasing intensity with increasing radius in the central regions before falling off beyond some break.  Type~II profiles are very abundant in local dwarf irregular galaxies (Herrmann et~al.\,2013; hereafter Paper~I) but these galaxies do not have spiral arms to cause radial migration, a method proposed to explain Type~II profiles in spirals, so are there similar radial color trends in dwarf galaxies as in spirals?  Disk-like dwarfs require a similar radial color analysis as that performed by BTP08 in spirals.

D.\,A.\,Hunter and collaborators have collected an extensive set of surface brightness profiles for 141 disk-like dwarf galaxies in up to 11 bands: H$\alpha$ \citep{he04} and $UBVJHK$ \citep{he06} from ground-based observations, 3.6~$\mu$m and 4.5~$\mu$m from $Spitzer$ \citep{hem06}, and FUV $+$ NUV from $GALEX$ \citep{hel10,deep}.  In Paper~I, we presented a human-assisted computerized method to fit the profiles of the full multi-wavelength data set and examined statistics of the fit parameters.

Here we focus on 319 {\it color} profiles in FUV$-$NUV (58), $U-B$ (129), and $B-V$ (132); the other colors ($J-H$, $H-K$, $3.6-4.5~\mu$m) have samples too small for analysis.  In Section~2 we overview the data sources.  Next we detail the classification and analysis of the color profiles.  Section~4 presents the radial color trend results and these are used in Section~5 along with mass-to-light ratios as a function of color (Herrmann et al., in preparation) to estimate radial mass profiles.  Finally we end with a discussion and some conclusions.  Future papers in this series will explore break relationships to \ion{H}{1} gas surface density and kinematics (in the smaller LITTLE THINGS\footnotemark[4] 41 galaxy subsample), and an examination of the two-dimensional images for each profile type.

\footnotetext[4]{The LITTLE (Local Irregulars That Trace Luminosity Extremes) THINGS (The HI Nearby Galaxy Survey) team is an international collaboration studying a multi-wavelength sample of 37 dIm and 4 BCD galaxies to understand star formation in these small systems \citep{LTdata}.}

\section{THE DATA}
Our galaxy sample (see Table~\ref{Sample}) is derived from the survey conducted by Hunter \& Elmegreen (2006) of nearby ($<$30 Mpc) late-type galaxies. The full survey includes 94 dwarf Irregulars (dIms), 26 Blue Compact Dwarfs (BCDs), and 20 Magellanic-type spirals (Sms). The 141~dwarf sample presented in the first paper of the present series (Paper~I) contains one fewer Sm galaxy and two additional dIm systems than the original survey. A multi-wavelength data set has been assembled for these galaxies. The data include H$\alpha$ images (129 galaxies with detections) to trace star formation over the past 10~Myrs (Hunter \& Elmegreen 2004) and satellite UV images (61 galaxies observed) obtained with the Galaxy Evolution Explorer ({\it GALEX}, Martin et al.\ 2005) to trace star formation over the past $\sim$200 Myrs (see, for example, Hunter et al.\ 2010, 2011; Zhang et al.\ 2012a). The {\it GALEX} data include images from two passbands with effective wavelengths of 1516 \AA\ (FUV) and 2267 \AA\ (NUV) and resolutions of 4\arcsec\ and 5.6\arcsec, respectively. Three of the galaxies in our sample with NUV data do not have FUV data. To trace older stars we have $UBV$ images, which are sensitive to stars formed over the past 1~Gyr for on-going star formation, and images in at least one band of $JHK$ for 40 galaxies in the sample, which integrates the star formation over the galaxy's lifetime. Note that nine dwarfs are missing $UB$ data and three more are missing $U$-band data. In addition we made use of 3.6~$\mu$m images (39 galaxies) obtained with the Infrared Array Camera (IRAC, Fazio et al.\ 2004) in the {\it Spitzer} archives also to probe old stars.

We were careful to remove or mask out foreground stars and background objects, and in most cases we made a 2D fit to the sky around the galaxy before subtraction. Significant care was taken to prevent over- or under-subtraction of the sky to ensure that breaks are not artifacts of data handling. The surface photometry was azimuthally averaged by integrating in fixed ellipses that increase in semi-major axis length in $\sim$10\arcsec\ steps. The geometric parameters (morphological center, position angle, and minor-to-major axis ratio $b/a$) were determined from the $V$-band outer isophotes and applied to all passbands the same way. The ellipse parameters were not allowed to vary with radius. See Hunter \& Elmegreen (2004, 2006) for more details on the surface photometry including the geometric parameters and ellipse widths. Finally, note that the {\it GALEX} photometry are on the AB magnitude system whereas the data in all other broad bands are on the Johnson/Cousin system.

\begin{figure*}
\epsscale{0.83}
\plotone{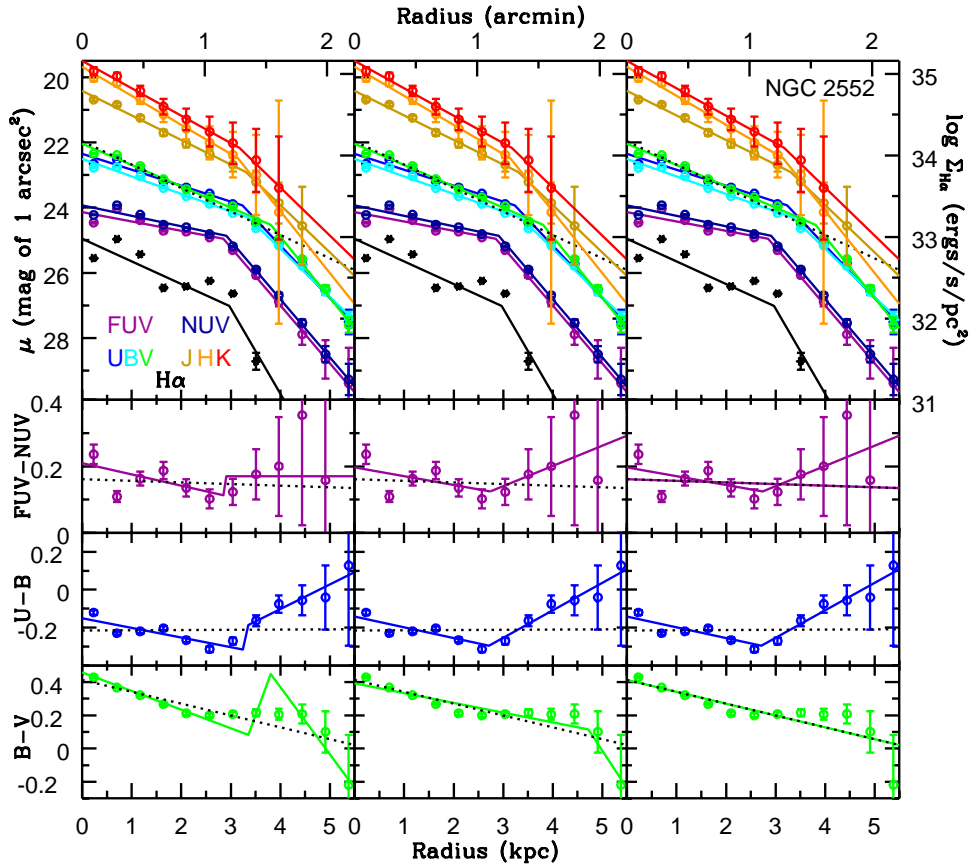}
\caption{Surface brightness (top row) and color profiles for an example galaxy, NGC~2552.  In each panel, the black dotted line shows the best linear fit to a full data set ($V$ band in the top row), weighted by $1/\sigma$.  The top row shows the surface brightness data and fits to all 9 passbands, including $V$.  The solid line fits to the {\it color} profiles indicate: the jagged subtraction of the surface brightness fits (left column), the best broken fit (middle column), and the final analyzed fits (right column).  Note that it is ambiguous if the FUV$-$NUV profile is better fit by a single line or two lines whereas the $U-B$ and $B-V$ profiles are clearly well fit by two lines and one line, respectively.  \label{n2552} }
\end{figure*}

For Paper~I, a program was written to determine (1) the best (single) fit to all the data of a given azimuthally averaged radial surface brightness profile and (2) the best way to fit the data with two lines (a broken fit).  (See the top row of Figure~\ref{n2552}.)  A minimizing $\chi^2$ function \citep{NumRec} was used to determine the best single line fit to all the radial data for the galaxy and passband under consideration.  These fits were weighted by $1/\sigma$ because the uncertainties are not distributed via a Gaussian function, instead ranging from 0.001 to $\sim$1 magnitude.  The $1/\sigma$ weighting was used as a compromise between no weighting at all and complete uncertainty weighting ($1/\sigma^2$), which overly weighted the bright inner data with low uncertainties.  The program also determined the best break location (i.e., lowest residual $\chi^2$ values with a break falling between the inner and outer sections) by the brute force method of breaking the radial profile into every possible pair of inner and outer subsets containing at least two data points, again using $1/\sigma$ weighting.  (See Section 3.2.4 in Paper~I about the complications from fitting only two data points.)  However, no weighting was used for the final broken fits.

The profiles of a few galaxies (9/141) were better fit with two breaks instead of just one and were given to the fitting program in two overlapping sections.  All broken profiles were classified as Type~III (outer slope shallower than the inner slope) or either Type~II or Type~FI (flat or increasing intensity with increasing radius before falling off beyond a break).  Empirically the FIs were assigned to profiles having an inner slope less than 0.075, corresponding to an inner scale length that is either greater than $\sim$14.5 kpc or negative due to the surface brightness increasing from the center to the break.  Note that 13/141 galaxies were fit with ``multiple'' profile types, such that profiles in different passbands have different profile types.  For many more details on the fitting process (also used in this study), see Section~3 of Paper~I.

\section{COLOR CLASSIFICATION AND ANALYSIS}
Our goal for this study was to explore whatever patterns exist in the radial color profiles of the 141 dwarf sample.  That is, looking from the center radially outward, do the colors get bluer (B), redder (R), or do they stay roughly flat (F)?  Furthermore, if there is a color break, then some combination of B, R, and/or F needs to be identified in as quantitative and unbiased a manner as possible.  There are eleven possible single or double trends: B, R, F, BB, BR (also referred to as ``U" shape), BF, RB, RR, RF, FB, FR.  (Note there is no FF because that would just be F.)  Once the trends have been identified, the radial positions of each profile can be scaled by the break location (arbitrarily defined in the redder band of the color) for Type~II and III profiles or by the scale length (also in the redder band) for Type~I.  Then the similar trends from the dwarfs can be averaged and compared and contrasted to the radial color trends found in spirals: BF for Type~Is, BR (``U'') for Type~IIs, and BRF (B+``S'' shape) for Type~IIIs.

While this process sounds simple enough, it is challenging in practice: unlike spirals where each profile type (I, II, or III) has a specific radial color trend that is seen in {\it each} galaxy not just the bulk average, dwarfs {\it do not}.  Perhaps this is not too surprising considering their irregular nature.  Type~II dwarfs in particular have color trends of various {\it flavors}\footnotemark[5], so classifying the radial color profiles was extremely tricky.  Additionally, large uncertainties on some colors, especially at radial distant points, further complicated the classification.  A critical question was: Is there really enough structure (i.e., disparity of individual data points from a single line fit) to merit a two piece fit, or does a single line agree with the data within or close enough to the error bars?  (See Figure~\ref{n2552} for some example color profiles.)  Furthermore, the elliptical radial bins for the surface photometry for each galaxy were separated by $\sim$10\arcsec\ to strike a balance between having low surface brightness uncertainties but a large enough number of radial bins to allow for a radial profile.  However, since the radial bin locations scaled by the break location or scale length for one galaxy are not necessarily anywhere near those scaled bin locations of another galaxy, some interpolation or fitting of the color profiles was critical for purposes of averaging colors from different dwarf galaxies.  The main goal in fitting the radial color trends is to be able to interpolate at any radius, as far out in radial extent as possible, especially for the color profiles of Type~II surface brightness dwarfs.  Fitting the color trends of Type~III profiles was not as important because the general horizontally stretched ``S'' shape (relatively constant colors, then steeply redder to the surface brightness break, then roughly constant colors beyond; see the rightmost column of Figures~1 and 2) is not captured well by a few lines, so natural cubic splines were used to include the nuances of the shape during the interpolation step instead of using one or two linear fits.

\footnotetext[5]{Since ``type'' is already used for galaxy classification (``Hubble type'') and Type~I, II, and III surface brightness profiles, we will use ``flavor'' (as used for quarks) to refer to different {\it color} profile trends.  Types~I, II, and III here refer {\it only} to surface brightness profiles.}

\begin{deluxetable*}{cccccccccc}
\tabletypesize{\scriptsize}
\tablecaption{Parameters of Single and/or Broken Fits to Color Profiles (Functions of Radius in kpc) \label{ColorFits}}
\tablewidth{0pt}
\tablehead{\colhead{Galaxy} & \colhead{Color} & \colhead{$R_{brk}$\tablenotemark{a}} & \colhead{CType\tablenotemark{b}} & \colhead{$m_I$\tablenotemark{c,d}} & \colhead{$b_I$\tablenotemark{d,e}} & \colhead{$m_i$\tablenotemark{c,d}} & \colhead{$b_i$\tablenotemark{d,e}} & \colhead{$m_o$\tablenotemark{c,d}} & \colhead{$b_o$\tablenotemark{d,e}}}
\startdata
0467-074 & $B-V$ & 6.13 & IIBF/B & -0.03$\pm$0.05 & 0.52$\pm$0.23 & -0.05$\pm$0.01 & 0.57$\pm$0.03 & 0.02$\pm$0.01 & 0.20$\pm$0.06\\
0467-074 & $U-B$ & 5.96 & IIF & 0.01$\pm$0.07 & -0.24$\pm$0.31 & \dots & \dots & \dots & \dots \\
1397-049 & $B-V$ & 4.47 & IIB & -0.05$\pm$0.09 & 0.70$\pm$0.32 & \dots & \dots & \dots & \dots \\
A1004+10 & $B-V$ & 1.70 & SI & 0.17$\pm$0.04 & 0.23$\pm$0.03 & 0.068$\pm$0.002 & 0.265$\pm$0.001 & 0.34$\pm$0.03 & 0.00$\pm$0.05\\
A2228+33 & $B-V$ & 2.69 & IIFR/F & 0.01$\pm$0.02 & 0.49$\pm$0.06 & 0.000$\pm$0.004 & 0.49$\pm$0.02 & 0.15$\pm$0.03 & -0.63$\pm$0.24\\
A2228+33 & $U-B$ & 2.57 & IIF & -0.01$\pm$0.02 & -0.05$\pm$0.08 & \dots & \dots & \dots & \dots \\
CVnIdwA & $B-V$ & 0.56 & IIBR & \dots & \dots & -0.30$\pm$0.23 & 0.21$\pm$0.08 & 1.45$\pm$0.52 & -0.74$\pm$0.37\\
CVnIdwA & $U-B$ & 0.52 & IIBR & \dots & \dots & -0.62$\pm$0.08 & -0.43$\pm$0.03 & 1.40$\pm$0.14 & -1.40$\pm$0.10\\
CVnIdwA & FUV-NUV & 0.42 & IIBR & \dots & \dots & -0.13$\pm$0.05 & -0.05$\pm$0.02 & 0.67$\pm$0.18 & -0.38$\pm$0.15
\enddata
\tablenotetext{}{Note: Table~2 is published in its entirety in the electronic edition of the {\it Astronomical Journal}.  A portion is shown here for guidance regarding its form and content.}
\tablenotetext{a}{Surface brightness profile break location (for Type~II and III profiles) or scale length (for Type~I profiles) in kpc.  For each color, the provided measurement is for the redder band ($V$ for $B-V$, $B$ for $U-B$, and NUV for FUV$-$NUV).}
\tablenotetext{b}{Color profile flavor, as determined in this study.  Again, the presence of a ``/'' indicates an ambiguous profile; the first classification was used in the broken-by-default analysis and the second in the single-by-default analysis.  The parameters for the best single and broken fits are listed for these ambiguous color trends.}
\tablenotetext{c}{Color slope, in mag arsec$^{-2}$ kpc$^{-1}$.}
\tablenotetext{d}{The subscripts indicate: $I$ = single line fit parameters whereas $i$ and $o$ = inner and outer fit parameters for broken fits, respectively.}
\tablenotetext{e}{The $y$ intercept for a color fit, in mag arsec$^{-2}$.}
\tablenotetext{}{Recall that the color profile fits for Type~III dwarfs with the characteristic horizontally stretched ``S'' color shape (SI or SB) were fit by natural cubic splines to capture the curvature in the color profile shape.  Consequently, the single and broken color fits were not very important in this analysis, but both are provided in the table.}
\end{deluxetable*}

Given the above complications, the classification method {\it started} with the subtraction of the fits of the surface brightness profiles as a visual guide.  In Paper I, although we found no statistical differences between the break locations in different bands, that does not mean that the profile in each band breaks at {\it exactly} the same radial distance.  For example, consider a galaxy (like NGC~2552) where the break in the $B$ profile ($R_{br,B}$) occurs closer in than the break in the $V$ profile ($R_{br,B} < R_{br,V}$).  The subtraction of the $V$ fit from the $B$ fit ($B-V$) is then $B_i - V_i$ (where $i$ indicates the inner components) out to $R_{br,B}$, then $B_o - V_i$ (where $o$ indicates the outer component) to $R_{br,V}$, followed by $B_o - V_o$ to the outer radial extent of the data.  Unfortunately, this subtraction of the fits to the $B$ and $V$ profiles results in some rather jagged $B-V$ ``fits'' (see the left column in Figure~\ref{n2552} for an example), and so it serves as a good $indication$ of the trends from the surface brightness profile fits, but the jagged fits are not adequate for averaging purposes.

The fitting program from Paper~I was adapted to fit the radial color trends under consideration here.  First all the color points corresponding to eliminated outlying surface brightness data (primarily due to deviations from star formation clumps, especially in the ultraviolet and $UBV$) in either of the two relevant bands for the given color were also eliminated from the radial color profile fitting.  Because the program was not sophisticated enough to determine if each color profile is best described by a single or broken fit, the program was used to determine the best {\it single} fit as well as the best {\it broken} fit for {\it each} color radial profile (see the middle column in Figure~\ref{n2552}).  A visual examination of these preliminary fits was used to identify 31 additional color points to be eliminated from 14 galaxies to aid in the final fitting; all but one of these data points were at the largest radii with large uncertainties and were removed from the analysis to allow the fits to be more reliable at interpolating data between the inner radial points.

The visual inspection was also used to determine which profiles are better fit with a single trend (B, R, or F) and which are better fit by a combination of two lines (see the right column in Figure~\ref{n2552}).  In making the decision about the best final fit (broken or single), the following were taken into account: (1) the similarity of the shape of the broken fit to that of the subtraction of the surface brightness fits (such as $B$ fit $-$ $V$ fit), (2) the agreement of the data to the single or broken fit, at least within error bars, (3) both segments of broken profiles must have at least two data points included per section, and (4) if the single and broken fits are very similar, the single fit was just used.  Happily, the colors of Type~I profiles were virtually always well fit with a single line and fitting the colors for the Type~III profiles was once again not critical because natural cubic splines were used for interpolation.  While there were many color profiles of Type~II dwarfs that were clearly well fit by a single line (112/230 = 49\%; 58/95 $B-V$, 41/95 $U-B$, and 13/40 FUV$-$NUV profiles) or two lines (54/230 = 23\%; 13/95 $B-V$, 27/95 $U-B$, and 14/40 FUV$-$NUV profiles) but not both, 28\% (64/230) of the color profiles of Type~II dwarfs (24/95 $B-V$, 27/95 $U-B$, and 13/40 FUV$-$NUV) were well fit by {\it both} a single line (within or close to the error bars) and two lines (with good agreement to the subtraction of the surface brightness fits or clearly passing through virtually every point).  Since the correct classification of these color profiles was so questionable, we performed {\it two} color analyses: one with all the ambiguous color profiles fit by {\it two lines} (``broken-by-default'') and one with all the ambiguous color profiles fit by a {\it single line} (``single-by-default'').

\begin{deluxetable*}{ccccccccc}
\tabletypesize{\scriptsize}
\tablecaption{Parameters of Fits to Average Profiles (Functions of Scaled\tablenotemark{a} Radii) \label{AveFits}}
\tablewidth{0pt}
\tablehead{\colhead{CType\tablenotemark{b}} & \colhead{Profile} & \colhead{Def?\tablenotemark{c}} & \colhead{$m_I$\tablenotemark{d,e}} &\colhead{$b_I$\tablenotemark{e,f}} & \colhead{$m_i$\tablenotemark{d,e}} & \colhead{$b_i$\tablenotemark{e,f}} & \colhead{$m_o$\tablenotemark{d,e}} & \colhead{$b_o$\tablenotemark{e,f}}}
\startdata
I(sp) & $B-V$ & \dots & \dots & \dots & -0.187$\pm$0.023 & 0.935$\pm$0.022 & -0.008$\pm$0.011 & 0.689$\pm$0.024 \\
IR & $B-V$ & \dots & 0.047$\pm$0.030 & 0.376$\pm$0.052 & \dots & \dots & \dots & \dots \\
II(sp) & $B-V$ & \dots & \dots & \dots & -0.221$\pm$0.015 & 0.858$\pm$0.009 & 0.149$\pm$0.022 & 0.531$\pm$0.026 \\
II(sp) & $B-V$ & 2B & -0.285$\pm$0.020 & 0.884$\pm$0.009 & -0.079$\pm$0.010 & 0.754$\pm$0.008 & 0.187$\pm$0.031 & 0.480$\pm$0.040 \\
IIB & $B-V$ & Sgl & -0.149$\pm$0.058 & 0.489$\pm$0.049 & \dots & \dots & \dots & \dots \\
IIB & $B-V$ & Brk & -0.152$\pm$0.063 & 0.479$\pm$0.053 & \dots & \dots & \dots & \dots \\
IIBF & $B-V$ & Brk & \dots & \dots & -0.199$\pm$0.002 & 0.536$\pm$0.001 & -0.018$\pm$0.003 & 0.403$\pm$0.004 \\
IIBR & $B-V$ & Sgl & \dots & \dots & -0.146$\pm$0.003 & 0.465$\pm$0.002 & 0.339$\pm$0.010 & 0.012$\pm$0.012 \\
IIBR & $B-V$ & Brk & \dots & \dots & -0.107$\pm$0.003 & 0.402$\pm$0.002 & 0.251$\pm$0.008 & 0.070$\pm$0.010 \\
IIF & $B-V$ & Sgl & -0.016$\pm$0.081 & 0.318$\pm$0.071 & \dots & \dots & \dots & \dots \\
IIF & $B-V$ & Brk & -0.016$\pm$0.086 & 0.305$\pm$0.075 & \dots & \dots & \dots & \dots \\
IIFR & $B-V$ & Brk & \dots & \dots & 0.007$\pm$0.001 & 0.273$\pm$0.001 & 0.110$\pm$0.004 & 0.168$\pm$0.005 \\
IIR & $B-V$ & Sgl & 0.082$\pm$0.059 & 0.251$\pm$0.050 & \dots & \dots & \dots & \dots \\
IIR & $B-V$ & Brk & 0.093$\pm$0.071 & 0.253$\pm$0.061 & \dots & \dots & \dots & \dots \\
III(sp) & $B-V$ & \dots & \dots & \dots & 0.128$\pm$0.015 & 0.665$\pm$0.011 & -0.284$\pm$0.089 & 1.150$\pm$0.120 \\
III(sp) & $B-V$ & 2B & -0.200$\pm$0.035 & 0.779$\pm$0.010 & 0.188$\pm$0.021 & 0.620$\pm$0.016 & -0.038$\pm$0.021 & 0.843$\pm$0.025 \\
III(SB) & $B-V$ & \dots & \dots & \dots & 0.347$\pm$0.011 & 0.221$\pm$0.006 & -0.005$\pm$0.004 & 0.580$\pm$0.005 \\
III(SB) & $B-V$ & 2B & 0.205$\pm$0.004 & 0.249$\pm$0.001 & 0.434$\pm$0.007 & 0.162$\pm$0.005 & 0.017$\pm$0.009 & 0.551$\pm$0.011 \\
III(SI) & $B-V$ & \dots & \dots & \dots & 0.321$\pm$0.006 & 0.211$\pm$0.003 & 0.044$\pm$0.009 & 0.456$\pm$0.010 \\
III(SI) & $B-V$ & 2B & 0.241$\pm$0.002 & 0.222$\pm$0.000 & 0.348$\pm$0.007 & 0.197$\pm$0.004 & 0.054$\pm$0.010 & 0.442$\pm$0.011
\enddata
\tablenotetext{}{Note: Table~3 is published in its entirety (including fits for $V$-band surface brightness ($\mu_V$), log of surface mass density ($\log \Sigma$), $U-B$, and FUV$-$NUV) in the electronic edition of the {\it Astronomical Journal}.  A portion is shown here for guidance regarding its form and content.}
\tablenotetext{a}{For broken and Type~I fits, the radii have been scaled by the surface brightness break location and scale length, respectively.}
\tablenotetext{b}{Color profile flavor, as determined in this study, for dwarfs as well as the color trends found in spirals from BTP08.}
\tablenotetext{c}{``Brk'' and ``Sgl'' indicate the broken-by-default and single-by-default samples, respectively. ``2B'' indicates a fit with two breaks; the parameters for the inner, middle, and outer segments are given in the $I$, $i$, and $o$ columns, respectively.}
\tablenotetext{d}{Slope, in mag arsec$^{-2}$ for a surface brightness or color fit or in $\log_{10} M_{\odot}$ pc$^{-2}$ for surface mass density.}
\tablenotetext{e}{The subscripts indicate: $I$ = single line fit parameters whereas $i$ and $o$ = inner and outer fit parameters for broken fits, respectively.}
\tablenotetext{f}{The $y$ intercept, in mag arsec$^{-2}$ for a surface brightness or color fit or in $\log_{10} M_{\odot}$ pc$^{-2}$ for surface mass density.}
\tablenotetext{}{Note that in some instances (II(sp), IIBR, III(SB), and III(SI)) log$\Sigma$ is fairly well fit by a single line.  The parameters for both the best single and broken fits are listed here in these four cases.  Additionally, only the two break (2B, but three segments) fits are shown in Figures~1 and 2 for the $B-V$, $U-B$, and FUV$-$NUV color profiles for Type~III BCDs and dIms (and spirals in $B-V$ only), but broken fits with only two segments are also listed here.}
\end{deluxetable*}

The program next classified each color profile according to the fitted trends.  Single profiles were either (1) flat~=~F, arbitrarily set by the absolute value of the slope being less than 0.02 mag arsec$^{-2}$ kpc$^{-1}$, (2) red~=~R, for positive slope, or (3) blue~=~B, for negative slope.  Each broken segment was also classified as B, R, or F, for the eight possible broken trends: BB, BR, BF, RB, RR, RF, FB, and FR.  The resulting fit parameters are provided in Table~2.  Of the three possible single trends and the eight possible broken trends, the following seven were found to be common in the broken-by-default analysis: B, R, F, BR, BF, RF, and FR whereas only a few color profiles were classified as BB, RB, RR, or FB.  All color profiles that were outliers from the common color trends are tagged (with an ``O'') in Table~1 and are discussed separately in Section 6.5.  In the single-by-default analysis, virtually all of the BF classifications became B or F and virtually all of the FR colors became R.  Though many ambiguous BR classifications became B, R, or F in the single-by-default analysis, large BR samples still remained for the averaging analysis.

As in BTP08, we wanted to average similar color radial profiles according to profile type, with the broken profiles normalized by the break location, to examine the average trends.  First we needed to isolate Type~II and III samples that extended a certain distance beyond the surface brightness break location, $R_{br}$, and Type~I samples that extended beyond some multiple of the disk light scale length, $h_R$.  We found that 1.5$R_{br}$ and 3$h_R$ yielded large enough samples while still probing a significant radial extent.  Using profiles that probed a shorter extent than these limits would have required extrapolation of fits in the averaging process.  All color profiles that were too short for the averaging analysis are tagged (with an ``S'') in Table~1.  Note that BTP08 normalized their spiral Type~I radii by 2.5 times the scale length, the typical break radius for spiral Type~IIs \citep{pt06}.  We instead normalized our dwarf Type~I profiles by the scale length since dwarf break locations are closer in than those of spirals.  We fit natural cubic splines to the color profiles for Type~III dwarfs and used the single or broken fits for Type~I and II profiles to interpolate at given normalized positions and then averaged these values according to trends and profile types.  The resulting average surface brightness and radial color profiles are shown in the top two rows of Figures~1 and 2 for $\mu_V$ and $B-V$ and the bottom two rows of Figures~1 and 2 for $U-B$ and FUV$-$NUV, where Figures~1 and 2 are for the broken and single by default samples, respectively.  The differences between these two samples will be discussed in Section 4.4.  Note that since FI profiles are just a special case of Type~II, all FI profiles were included with the Type~II class in the averaging process.  The parameters for the fits to the average profiles are provided in Table~3.

The nine two break and thirteen multiple type galaxies require some additional explanation.  For isolating the radial color trends for Types~I, II, and III, the profile type in FUV, $U$, or $V$ was used for the multiple type galaxies for FUV$-$NUV, $U-B$, and $B-V$, respectively.  Most of the two break galaxies have color trends that are only common flavors for the second break but outliers for the first break.  DDO~126 and Mrk~178 have common color trends for both breaks; the first break was used for both of these dwarfs for the averaging analysis because scaling the radius by the second break location would have yielded normalized profiles that would have been too short for the averaging analysis.  DDO~53 and Haro~43 were normalized by the first or second break based on the color.

\begin{figure*}
\epsscale{0.91}
\plotone{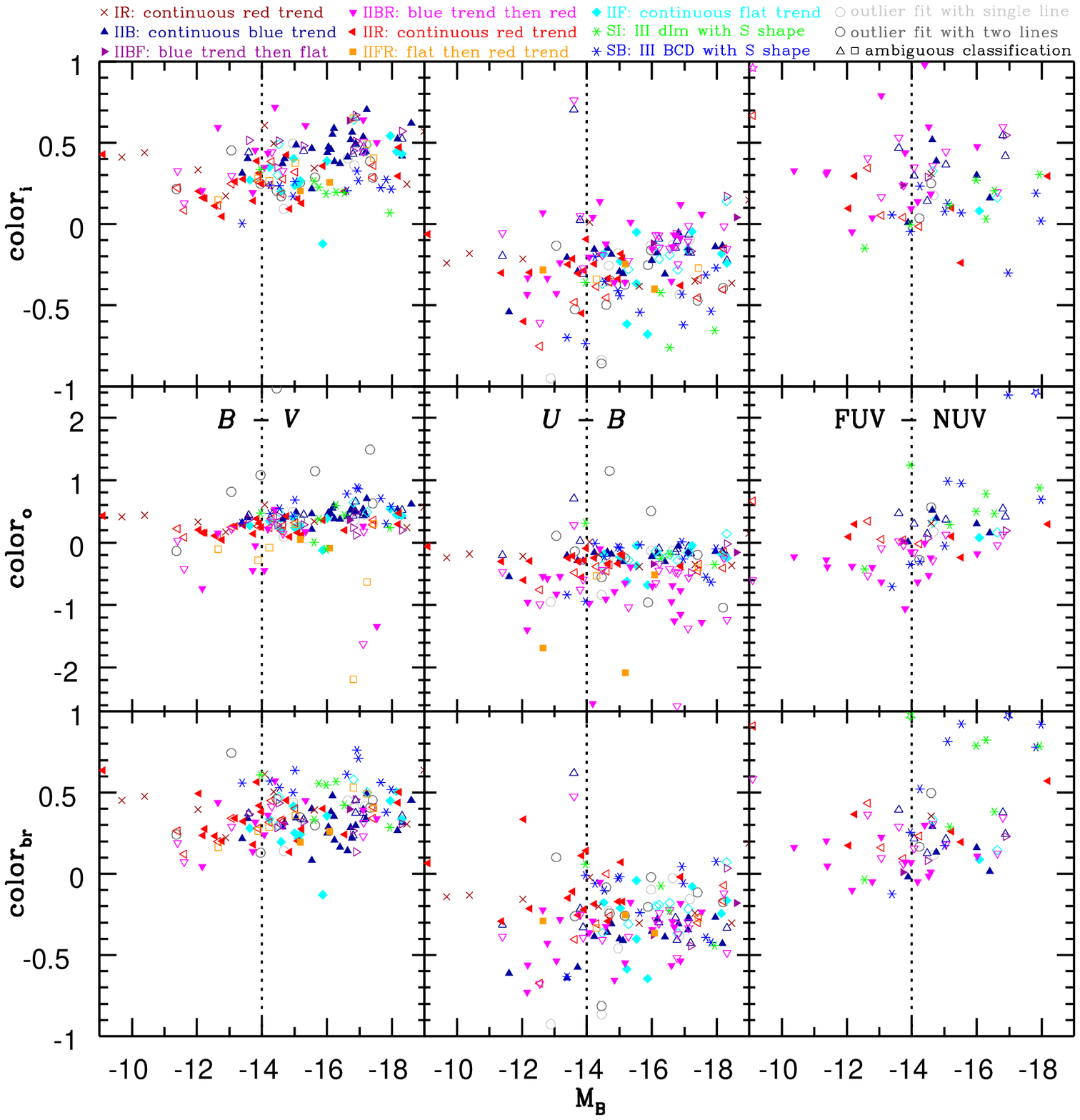}
\caption{Three colors (in mag arcsec$^{-2}$) at three specific locations: ({\it top}) projected central value from the inner fit, color$_i$, ({\it middle}) projected central value from the outer fit, color$_o$, and ({\it bottom}) value at the break, color$_{br}$ or one scale length (for single fit colors).  Since B, R, and F color trends have no second component, their inner values are plotted in both the top and middle panels for comparison.  Profiles with ambiguous classifications, indicated by open symbols of the relevant color and shape, are plotted twice per panel but always directly vertical of each other.  Any star across the top or bottom of a panel points toward a data point that lies outside the $y$ range of the plot.  Note a slight trend with $M_B$ especially between IIR and IIB points in $(B-V)_i$ but the lack of this trend in $(B-V)_{br}$.  Also note that Type~III SI and SB colors tend to be bluer in color$_i$ but redder in color$_o$ and color$_{br}$ than Type~II colors. \label{colors} }
\end{figure*}

\begin{figure*}
\epsscale{0.91}
\plotone{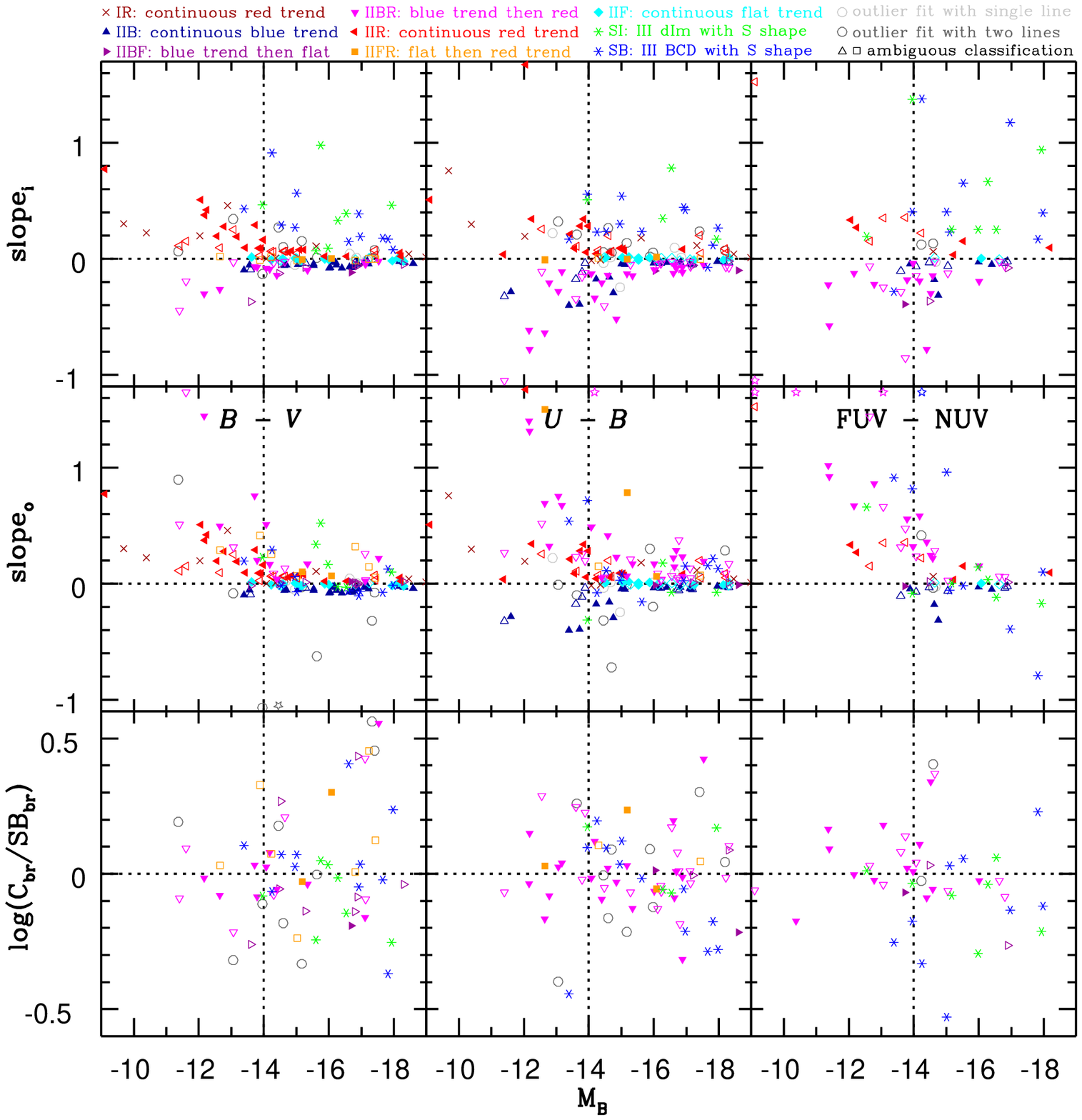}
\caption{Inner slope, slope$_i$ ({\it top}) and outer slope, slope$_o$ ({\it middle}) both in mag arcsec$^{-2}$ kpc$^{-1}$ and log$_{10}$ ratio of the color break location to the surface brightness break location (C$_{br}/$SB$_{br}$) ({\it bottom}) for $B-V$, $U-B$, and FUV$-$NUV as a function of $M_B$.  Again, since B, R, and F color trends have no second component, their single slopes are plotted in both the top and middle panels for comparison.  See the caption of Figure~\ref{colors} for more information.  Note that fainter Type~II dwarfs tend to have steeper slopes than their brighter counterparts and that Type~III dwarfs have red inner slopes.  Also note that the break location in color profiles is roughly correlated with the surface brightness break location (C$_{br}/$SB$_{br}\sim 1$), though with significant scatter. \label{slopes} }
\end{figure*}

\section{COLOR FIT RESULTS}

\subsection{Fit Parameter Results: Colors}
The top row of Figure~\ref{colors} shows the projected central color from the inner fits of the $B-V$, $U-B$, and FUV$-$NUV color profiles as functions of $M_B$.  Generally, only brighter ($-14 > M_B > -19$) Type~II dwarfs have continuously blue (IIB navy $\bigtriangleup$) or flat (IIF cyan $\Diamond$) trends leaving fainter ($-9 > M_B > -14$) Type~II dwarfs with almost solely red (IIR red $\triangleleft$) or blue then red (IIBR, or ``U'' magenta $\bigtriangledown$) color trends, especially in $B-V$ (also see Section 4.4).  The relatively bright Sm subsample contributes to the large number of bright IIB dwarfs (see Section 6.3).  Furthermore, there is a slight trend that the central color tends to be bluer in fainter Type~II dwarfs and redder in brighter Type~II dwarfs.

The middle row of Figure~\ref{colors} shows the projected central color from the outer fits where the central color for color profiles well fit by a single line are shown again here for sake of comparison.  The IIBR (magenta $\bigtriangledown$) points are generally on the red side of the IIR (red $\triangleleft$) and IIB (navy $\bigtriangleup$) points in color$_i$ but on the blue side in color$_o$, which makes sense considering the directions of the slopes.

The bottom row shows the color at the color break location (for profiles well fit by two lines) or at one surface brightness scale length (for profiles well fit by a single line).  The general trend with $M_B$ from color$_i$ (especially in $B-V$) disappears in color$_{br}$.  Even though only the brighter Type~II dwarfs tend to have IIB or IIF trends and have slighter redder centers than dimmer Type~II dwarfs generally with IIR or IIBR trends, at the color break location or one scale length the $B-V$ and $U-B$ colors are roughly the same with $M_B$, though with considerable scatter.

The Type~III dwarfs (green SI for dIms and blue SB for BCDs, both * in the figure) are virtually all in the brighter half of the full sample.  Furthermore, the Type~III colors tend to be bluer in color$_i$ but redder in color$_o$ and color$_{br}$ than Type~II dwarf colors.  This is a consequence of the general horizontally stretched ``S'' shaped curve of Type~III color profiles, evidently still caught in the two line fit even though natural cubic splines were used for the averaging analysis to capture and interpolate the ``S'' more accurately. 

\subsection{Fit Parameter Results: Slopes and Breaks}
The top two rows of Figure~\ref{slopes} show the inner and outer slopes of the $B-V$, $U-B$, and FUV$-$NUV color profiles, respectively, again as functions of $M_B$.  Type~III dwarfs (SI for dIms and SB for BCDs) clearly have red inner slopes in all three colors.  Also, fainter ($-9 > M_B > -14$) Type~II dwarfs (probably because they are smaller) tend to have steeper slopes than brighter ($-14 > M_B > -19$) Type~II dwarfs, especially the continuously red IIR (red $\triangleleft$) slopes in $B-V$ and $U-B$ and both the inner and outer slopes of ``U''-shaped IIBR (magenta $\bigtriangledown$) color profiles in all three colors.  The bottom row of Figure~\ref{slopes} presents the ratio of the color profile break location to the surface brightness break location (from the redder band of the color), log(C$_{br}/$SB$_{br})$, and reveals no trend with $M_B$, but significant scatter around C$_{br}/$SB$_{br}\sim 1$.  Even in Type~III color profiles the color break frequently occurs near the surface brightness break, which was not necessarily expected considering the more complicated stretched ``S'' shaped color profiles of Type~IIIs.  Evidently the two line fits generally captured the middle red trend and the outermost flattish color trend just beyond the surface brightness break, consequently missing the inner flattish trend.

\subsection{Type I Radial Color Trends}
The leftmost columns of Figures~1 and 2 show radial color trends from dwarfs with Type~I surface brightness profiles and are identical since no color profile for Type~I dwarfs had an ambiguous classification.  The colors of a few Type~I dwarfs are relatively flat with radius (IF), but typically there is a slight reddening (IR, indicated in red) trend with increasing radius.  Since there was a small number of dwarfs with Type~I surface brightness profiles, all the Type~I profiles with flat and continuously red colors were averaged together.  Only one dwarf galaxy (DDO~99) was found to have a Type~I profile in the ultraviolet so the fit is shown, but obviously no average was possible.  The gold profiles in Figures~1 and 2 represent the averages for spirals from BTP08.  The $g'$, $r'$, and $g'-r'$ data were digitized and transformed into $V$ and $B-V$ using $V = g' - 0.55(g'-r') - 0.03$ and $B-V = 1.02(g'-r')+0.20$ from \citet{SDSSfilters}.

\begin{figure}
\epsscale{0.87}
\plotone{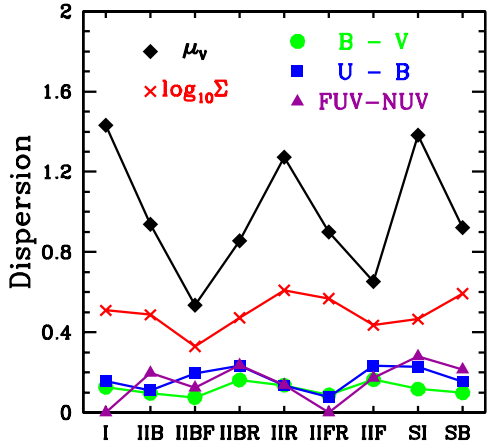}
\caption{The average dispersion around the mean trend in various quantities ($\mu_V$, $\log_{10} \Sigma$ where $\Sigma$ is the surface mass density discussed in Section 5, $B-V$, $U-B$, and FUV$-$NUV) as a function of the common color profile flavors for the broken-by-default sample analysis. \label{dispersion} }
\end{figure}

\subsection{Type II Radial Color Trends}
The middle two columns of Figures~1 and 2 show the variety of radial color trends seen in Type~II dwarfs for the broken and single by default samples, respectively.  There are few differences between the two samples; the most obvious is the disappearance of the IIBF (blue then flat) and IIFR (flat then red) categories in the single-by-default results due to the previously discussed transition of virtually all the broken-by-default IIBF profiles to either IIB (continuously blue) or IIF (flat) and virtually all the IIFR profiles to IIR (continuously red) in the single-by-default samples.  Otherwise, with respect to the broken-by-default average results, the single-by-default (1) $B-V$ IIBR (or ``U'') average color profile shifted to the red by $\sim$0.1~mag~arcsec$^{-2}$ on both ends, (2) $U-B$ IIB average color profile shifted to the red by $\sim$0.1~mag~arcsec$^{-2}$, (3) $U-B$ IIR average color profile shifted to the blue by roughly $\sim$0.1~mag~arcsec$^{-2}$, (4) FUV$-$NUV IIBR average color profile shifted to the blue by 0.05-0.1~mag~arcsec$^{-2}$, and (5) outer end of the FUV$-$NUV IIR average color profile shifted to the red by $\sim$0.1~mag~arcsec$^{-2}$.  The remaining average Type~II color profiles varied very little between the two samples.

Regardless of the sample, an interesting observation is that Type~II dwarfs with red trends (IIR) have a wider distribution of $V$ band surface brightness from overall faint to overall bright profiles than any other Type~II color classification.  Figure~\ref{dispersion} shows the dispersion in various quantities ($\mu_V$, $\log_{10} \Sigma$ where $\Sigma$ is the surface mass density discussed in Section 5, $B-V$, $U-B$, and FUV$-$NUV) as a function of the various color profile flavors and reveals the larger dispersion in $\mu_V$ in IIR than other Type~II color classification.

\begin{figure*}
\epsscale{1.0}
\plotone{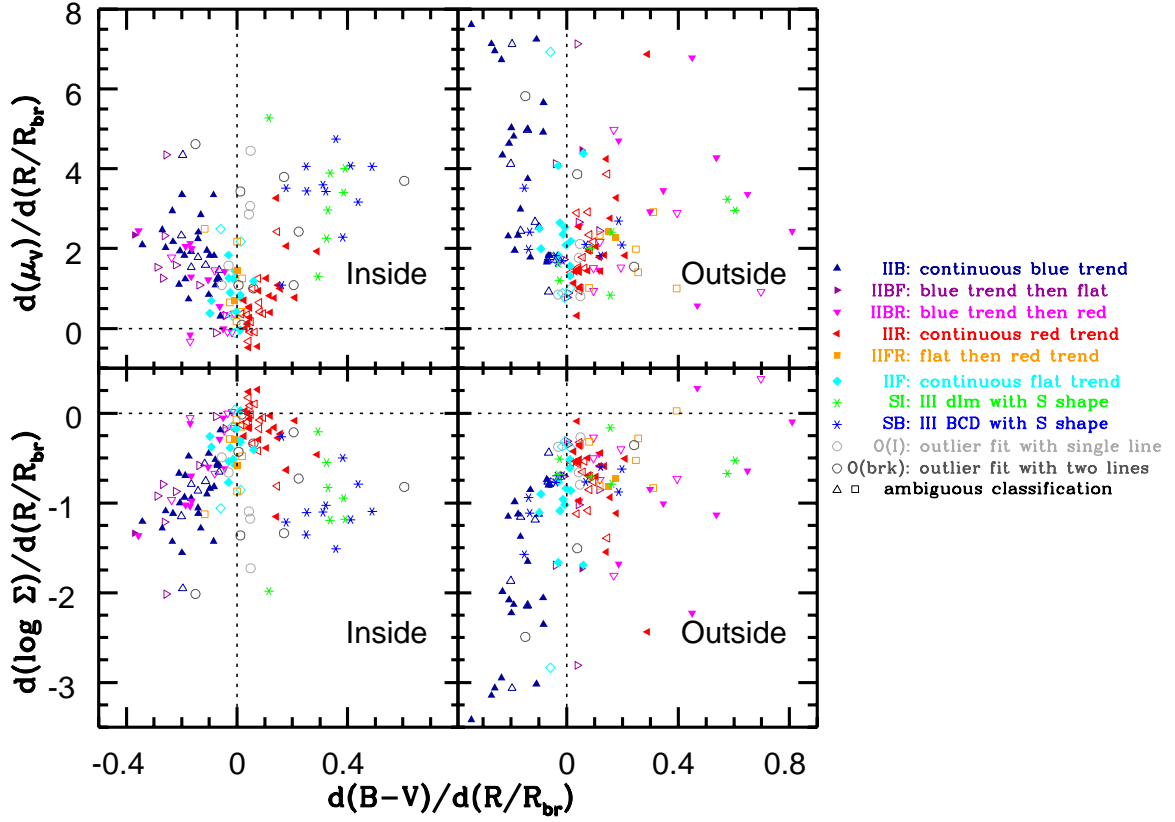}
\caption{Slopes of $V$ band surface brightness ($\mu_V$, {\it top}) and $\log_{10}$ of surface mass density ($\Sigma$, {\it bottom}) both with respect to the radius scaled by the break location ($R_{br}$) as functions of the slope of $B-V$ with respect to scaled radii.  The left and right panels show the parameters for inside and outside of the surface brightness $R_{br}$, respectively.  For continuous color trends (IIB, IIR, and IIF), the $B-V$ slope does not change from the inside to the outside.  Note the continuous trend from IIB/IIBF/IIBR through IIF/IIFR to IIR points in the {\it inside} panels for both $\mu_V$ and $\log_{10} \Sigma$ whereas the Type~III points are highly separated out to the red. \label{muSigslopes} }
\end{figure*}

Another result is that Type~II dwarfs with flat (IIF) or red (IIR) color trends tend, on average, to have shallower inner $V$ band surface brightness slopes than any kind of Type~II dwarf with some blue trend (IIB, IIBR, or IIBF).  This can be seen in the top left panel of Figure~\ref{muSigslopes} showing the slope of the inner $\mu_V$ with respect to the radius scaled by the surface brightness break location ($R/R_{br}$) plotted as a function of the inner $B-V$ color slope, also with respect to $R/R_{br}$.  Note that since we fit both $\mu_V$ and $B-V$ as linear functions of radius (in kpc), $d(\mu_V)/d(R/R_{br}) = \mu_{slope} \times R_{br}$ and $d(B-V)/d(R/R_{br})=(B-V)_{slope} \times R_{br}$.  Interestingly, there is a continuous trend from steeper inner $\mu_V$ slopes in IIB, IIBF, and IIBR profiles through shallower slopes in IIF and IIFR profiles to even shallower slopes in IIR profiles.  However, the Type~III data are separate from the Type~II points with steeper inner $\mu_V$ slopes as well as significantly steeper red $B-V$ color profiles.  This is a further indication that the {\it insides} of dwarf galaxies are drastically different between Type~II and III galaxies as we previously noted in Paper~I of this series.  The top right panel of Figure~\ref{muSigslopes} shows the slope of the outer $\mu_V$ profiles versus the outer $B-V$ slope.  Here the Type~III data have collapsed onto the Type~II points.  It is not surprising that $d(\mu_V)/d(R/R_{br})$ is steeper in the outside than the inside of Type~II dwarfs whereas the reverse is true for Type~III dwarfs since this is the definition of Type~II and III surface brightness profiles.

\begin{figure*}
\epsscale{0.9}
\plotone{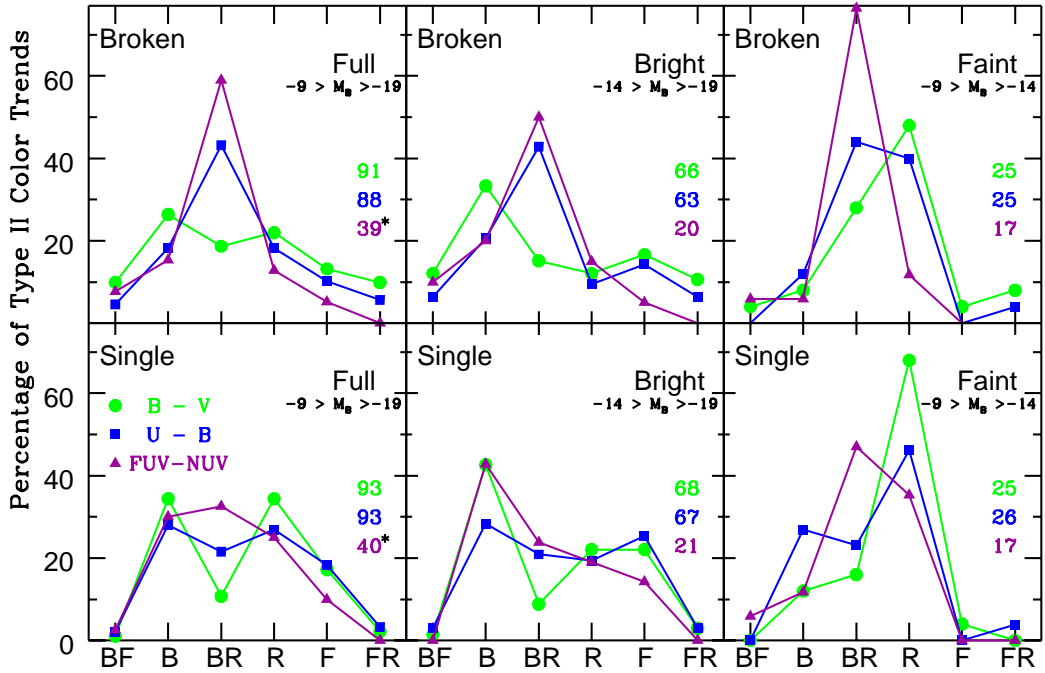}
\caption{The percentage of the different Type~II color trends.  In each case, the percentage is with respect to the total number of Type~II profiles of the six flavors shown (provided as the colored numbers at the right of each panel), including profiles that were too short for the averaging analysis but excluding Type~II profiles tagged as being outliers.  ``Broken'' and ``Single'' indicate the results from the broken and single by default analyses, respectively.  The break down of the full Type~II sample is shown in the left column whereas the bright ($-14 > M_B > -19$) and faint ($-9 > M_B > -14$) subsets are shown in the central and right columns, respectively. *: Two of the galaxies with FUV$-$NUV colors did not have $B$ data, so could not be separated into the bright or faint subsets. \label{percent} }
\end{figure*}

The main difference between the results of the broken and single by default analyses is the distribution of the different Type~II color flavors as shown by Figure~\ref{percent}.  In the full broken-by-default analysis, the largest percentage of $U-B$ and FUV$-$NUV Type~II profiles have a IIBR (``U'') color shape, like Type~II spirals, but the next most common Type~II color flavors are IIB and IIR whereas IIB and IIR are a bit more common than IIBR in $B-V$.  However, in the full single-by-default analysis, the IIBR dominance over the IIB and IIR flavors is reduced since many ambiguous IIBR classifications became IIB, IIR, or IIF in the single-by-default analysis.  When the full sample of Type~II color trends (without outliers) are separated into two $M_B$ bins (``bright'': $-14 > M_B > -19$ and ``faint'': $-9 > M_B > -14$), virtually all the ``faint'' Type~II dwarfs are IIR or IIBR whereas the ``bright'' Type~II dwarfs additionally have significant populations of IIB and IIF flavors.  Regardless of the analysis sample, IIBF and IIFR flavors are generally rare.

\subsection{Type III Radial Color Trend}
The color profiles of dwarf Type~IIIs have a general horizontally stretched ``S'' shape such that the colors are initially relatively constant (or at least only very gradually getting redder), then become fairly steeply redder to the surface brightness break location, and then remain roughly constant again (see the rightmost column of Figures~1 and 2).  As with the dwarf Type~Is, the Type~III columns in Figures~1 and 2 are identical because none of the color profiles for Type~III surface brightness profiles had ambiguous classifications.  The two identical outer columns hopefully aid in comparison of the different inner columns between Figures~1 and 2.  Originally the program classified the color trends of Type~III dwarfs primarily as R, RF, RB, or RR but visually these were reclassified as SB or SI for BCDs and dIms, respectively, to capture the full flavor of the complex trend where ``S'' = S shape.  Clearly there is a slightly stronger gradient in BCDs than in dIms, at least in $U-B$ and FUV$-$NUV.

\section{ESTIMATED STELLAR MASS PROFILES}

BTP08 noted that it is straightforward to determine the stellar surface mass density, $\Sigma$, profile from the surface brightness, $\mu$, and color profiles using
\begin{equation}
\log_{10} \Sigma = \log_{10} (M/L)_{\lambda} - 0.4(\mu_{\lambda}-m_{\rm{abs},\odot,\lambda}) + 8.629,
\end{equation}

\noindent where $m_{\rm{abs},\odot,\lambda}$ is the absolute magnitude of the Sun at wavelength $\lambda$ and $\Sigma$ is measured in $M_{\odot}$~pc$^{-2}$, but only if a reliable relationship between some color and the mass-to-light ratio $M/L$ is known.  For their spirals, BTP08 used the following expression:
\begin{equation}
\log_{10} (M/L)_{\lambda} = (a_{\lambda} + b_{\lambda} \times \rm{color}) - 0.15,
\end{equation}

\noindent where the 0.15 factor was used for a Kroupa initial mass function (IMF) (Kroupa 2001) and for $\log_{10}(M/L_{r'})$ vs.\,$g'-r'$, $a_{\lambda} = -0.306$, and $b_{\lambda} = 1.097$ \citep{b+03}.  Since it is questionable if this relationship is suitable for dwarf galaxies, with lower metallicities and potentially different star formation histories than spirals, we determined a new correlation between $\log(M/L)$ and various color for dwarfs. Those relationships are presented elsewhere (Herrmann et al., in preparation).

We used the $a_V$ and $b_V$ values from Herrmann et al.\ (in preparation), the $V$-band surface brightness and $B-V$ profiles, and a solar absolute magnitude in $V$ of 4.80 (http://mips.as.arizona.edu/$\sim$cnaw/sun.html) to determine the stellar mass density profiles for the remaining galaxies in our sample of 141 dwarfs.  (Actually, 9/141 galaxies are missing $B$-band data, so the total $B-V$ sample is 132.)  Note that mass profiles were only calculated for the $B-V$ data set; the mass-to-light ratio is not correlated as strongly with $U-B$ as with $B-V$.  This makes sense because redder passbands trace the mass better than bluer passbands.  The FUV$-$NUV colors would not have been useful at all for estimating $M/L$.

We applied an inclination correction by using the observed $b/a$ values from the $V$-band morphology of the outermost isophote by \citet{he06} and an assumed intrinsic $b/a_0$ value of 0.3 determined by \citet{hh66,vdB88}.  As a check, for the 34 galaxies from the sample of \citet{z+12a}, for which spectral energy distribution-determined surface mass density profiles were available, we compared the inclination corrected mass profile radial values to those determined from our analysis and found excellent agreement.  However, for the averaging analysis we used the mass profiles from \citet{z+12a} for those 34 galaxies.

\subsection{Stellar Surface Mass Density Profile Results}
The third rows in Figures~1 and 2 show the resulting radially normalized and averaged surface mass density profiles for the broken and single by default analyses, respectively.  In both cases, the average surface mass density profile resulting from the dwarfs with Type~II surface brightness profiles with BR (``U'' shaped) color profiles is relatively well fit by a single line, as BTP08 found with Type~II spirals.  That is, the break in the surface brightness profiles for Type~IIBR galaxies (spirals and dwarfs) is greatly reduced in the surface mass density profiles.  However, the break {\it remains} for the other Type~II dwarfs with different color flavors (IIB, IIF, IIR...).  Surprisingly, the break is also greatly {\it reduced} in the surface mass density profiles of dwarf Type~III SI (dIms) and SB (BCDs) even though the Type~III break clearly remains in the surface mass density profiles of spirals.  (See Section 6.2.)

The decreased break in the surface mass density relative to the surface brightness for the IIBR, SI, and SB dwarfs can also be seen in the lower panels of Figure~\ref{muSigslopes} which plot the slope of the surface mass density profile against the slope of the $B-V$ trend.  The magenta IIBR, green SI, and blue SB points have roughly the same surface mass density slopes ($d(\log\Sigma)/d(R/R_{br})$) in both the inside and outside, so those data points generally shift horizontally (due to having different $B-V$ slopes) but do not shift by much in the vertical direction.  Note that combining equations 1 and 2 from Section 5 with our linear fits of surface brightness ($\mu$) and colors (specifically $B-V$) yields
\begin{equation}
\frac{d(\log \Sigma)}{d(R/R_{br})} = R_{br} \times [b_{\lambda} (B-V)_{slope} - 0.4 \mu_{slope}]
\end{equation}
\noindent where $b_V = 1.081$.  Considering the dependence on $-0.4 \mu_{slope}$, it is not surprising that the shallowing trend from IIB/IIBF/IIBR through IIF/IIFR to IIR points in inner surface brightness slope as a function of $B-V$ slope seen in Figure~\ref{muSigslopes} (top left) also appears in the plot of inner $\log \Sigma$ slope (bottom left).  However, the consequence of the $b_{\lambda} (B-V)_{slope}$ term is to rotate all the $d(\log \Sigma)/d(R/R_{br})$ points slightly counterclockwise.  That is, applying the color slope term to a pure calculation of $-0.4 \mu_{slope} R_{br}$ slides the blue data slightly downward and the red data slightly upward.

\begin{figure}
\epsscale{1.1}
\plotone{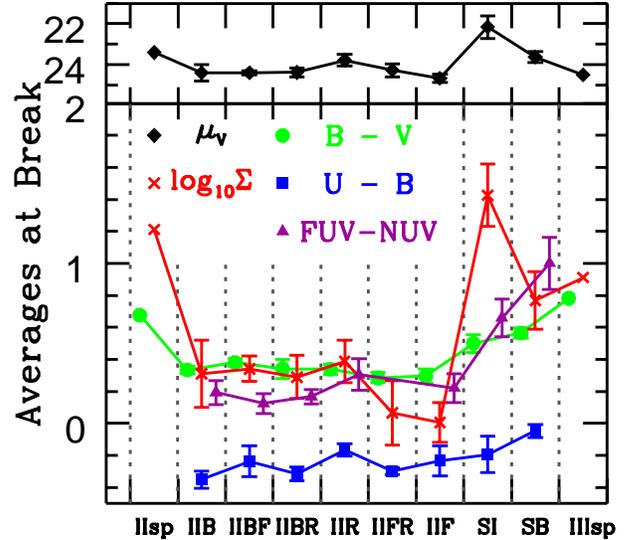}
\caption{The averages of various quantities ($\mu_V$, $\log_{10} \Sigma$, $B-V$, $U-B$, and FUV$-$NUV) evaluated at the surface brightness break location as a function of the common color profile flavors. The values are the averages between the broken-by-default and single-by-default sample analyses, though the differences are small.  The error bars show the uncertainty in the mean (as in Figures 1 and 2) and are the larger value between the two sample analyses. The $B-V$ and FUV$-$NUV points have been shifted slightly to the left and right, respectively, for ease of viewing and to prevent error bars from overlapping. \label{averages} }
\end{figure}

\begin{figure*}
\epsscale{1.1}
\plottwo{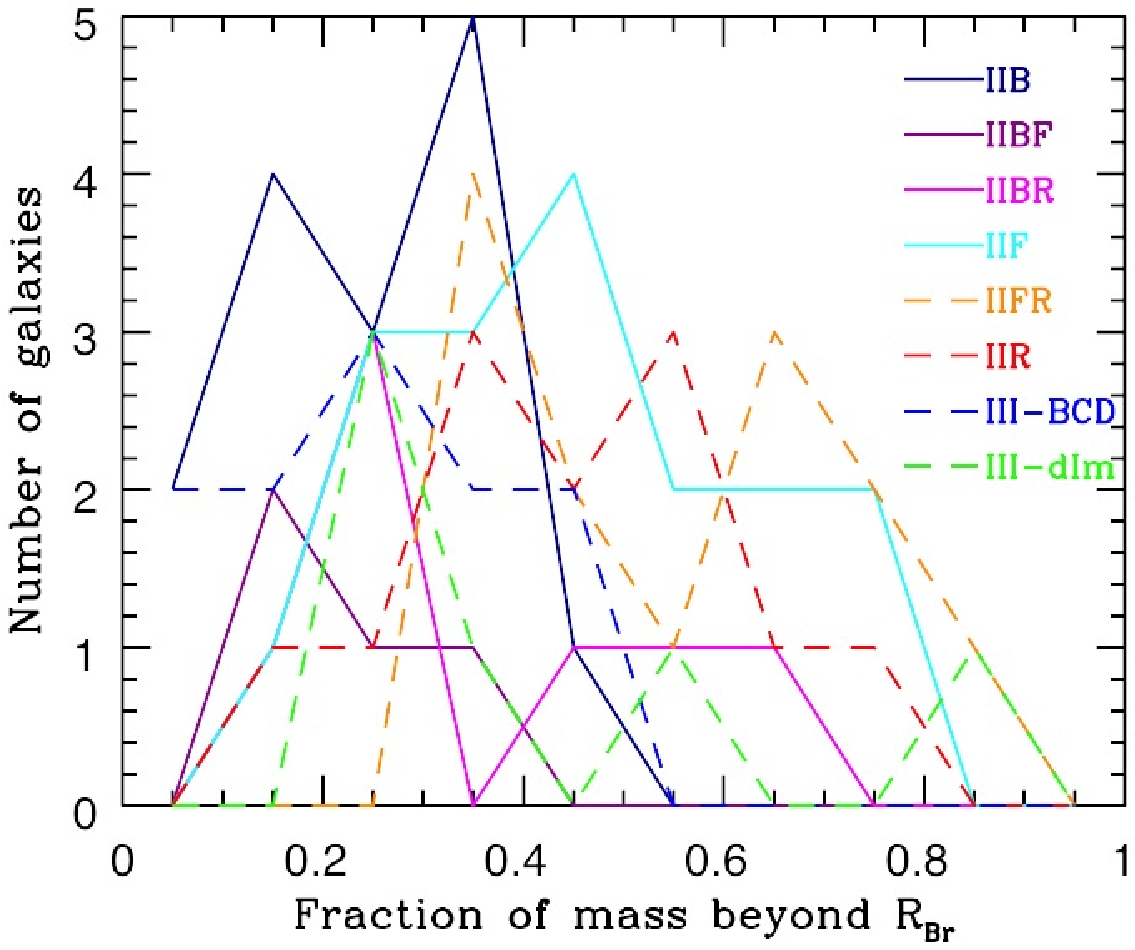}{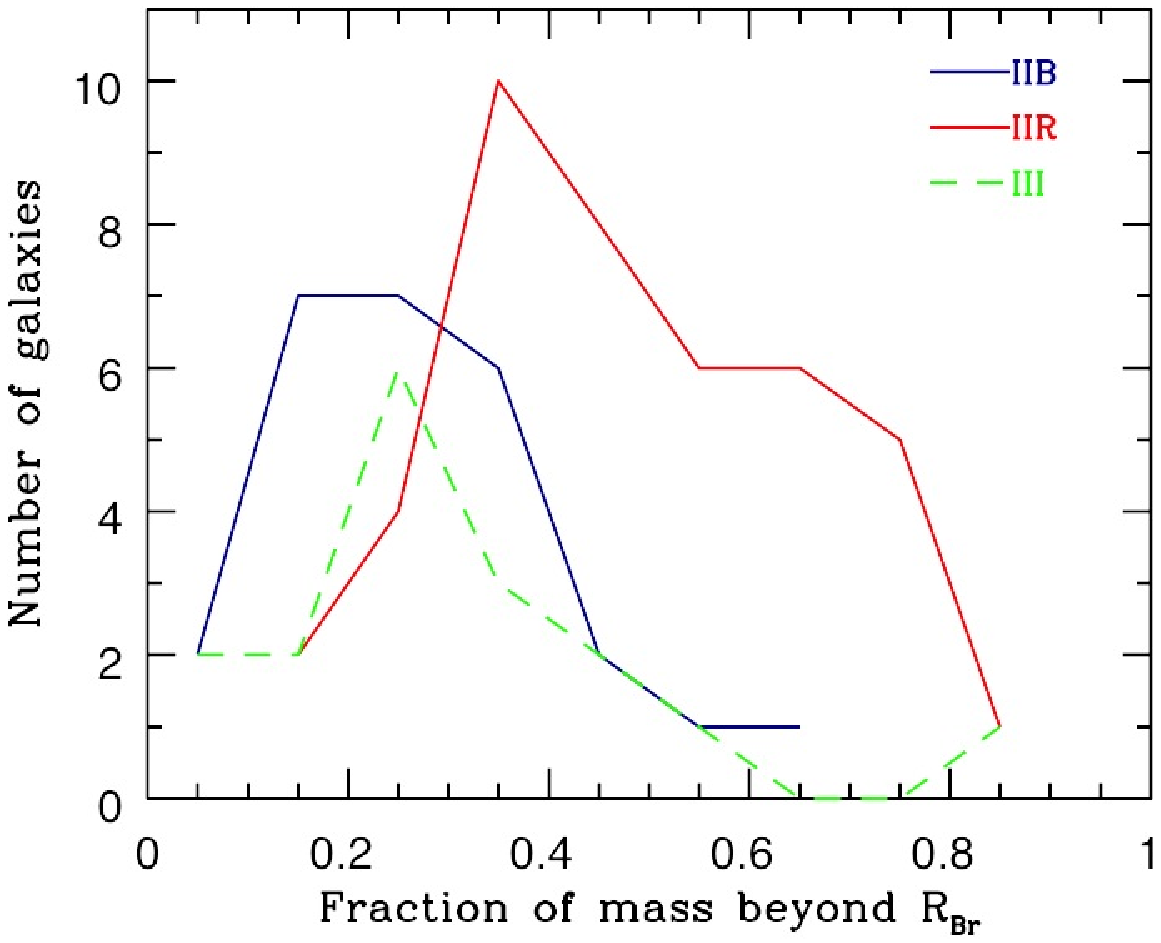}
\caption{Number of galaxies with the given fraction of the stellar mass beyond the break in the $V$-band surface brightness profile. Galaxies are classified by color profile flavor given in Table 2.  The color coding of the groups is the same as that in Figures 4 and 5.  \label{fig-mass}}
\end{figure*}

\subsection{Stellar Surface Mass Density at the Break}
Figure~\ref{averages} provides the averages of various quantities (after being scaled by the surface brightness break location) evaluated at the surface brightness break locations ($R/R_{br} = 1$).  The surface mass density at the break is roughly 2.2 $M_{\odot}$~pc$^{-2}$ in Type~II dwarfs with an inner blue trend (IIB/IIBF/IIBR) or a continuous red trend (IIR) and a bit lower (1.1 $M_{\odot}$~pc$^{-2}$) for Type~II dwarfs with flat (IIF) or flat then red (IIFR) trends.  These are all lower than the $13.6 \pm 1.6$~$M_{\odot}$~pc$^{-2}$ that BTP08 found for Type~II spirals.  However, the surface mass density at the break in Type~III BCDs with the characteristic ``S'' shape (SB) of 5.9~$M_{\odot}$~pc$^{-2}$ is not too far from that of BTP08 Type~III spirals ($9.9 \pm 1.3$~$M_{\odot}$~pc$^{-2}$) whereas $\Sigma_{br}$ is larger (27~$M_{\odot}$~pc$^{-2}$) in Type~III dIms (SI).  Note also in Figure~\ref{averages} that $\mu_V$ at the break location is roughly 24 mag~arcsec$^{-2}$ for all the Type~II and III galaxies considered here as we found in Paper~I, except our small sample of Type~III dIms break slightly brighter ($\sim$22 mag~arcsec$^{-2}$), and the colors at the breaks in Type~II dwarfs are roughly constant ($B-V \sim 0.3$, $U-B \sim -0.3$, and FUV$-$NUV $\sim 0.2$) and slightly bluer than Type~II spirals and Type~III dwarfs and spirals.  From Starburst99 \citep{SB99a,SB99b,SB99c,SB99d}, a $B-V$ color of $\sim$0.3 corresponds to an age of 10 Gyr for continuous star formation at roughly 1/5 solar metallicity.

\subsection{Stellar Mass Fraction beyond the Break}
We have determined the fraction of the stellar mass that has formed beyond the surface brightness break as measured from the $V$-band profile.  We determined the masses by integrating the surface mass density profiles.  The galaxies are classified according to the $B-V$ color profile flavor given in Table 2 (broken-by-default), and the histogram is given in Table \ref{tab-mass} and shown in Figure \ref{fig-mass}.  Galaxies with Type~III surface brightness breaks are put into two groups: BCDs (III-B) and dIms (III-I).  The left panel includes all of the groups given in the Table, and the color coding of the groups is the same as that in Figures 1, 2, 4, 5, and 7.  In the right panel IIB (blue) includes IIB (continuously blue), IIBF (blue then flat), and IIBR (blue then red); IIR (red) includes IIF (flat), IIFR (flat then red), and IIR (continuously red); and III includes III-BCD and III-dIm.

We see that the color flavors that have a significant number of galaxies with more than 50\% of the mass beyond the break radius include those with steeper outer profiles that become redder with radius (IIR and IIFR).  Type~II profiles with flat color profiles (IIF) span the range from 20\% to 80\% stellar mass beyond the break, while the Type~II profiles with blue inner color profiles (IIB, IIBF, and IIBR) and the Type~IIIs tend to have less than 40\% of their total stellar mass beyond the break radius.  Thus, galaxies with a significant stellar mass beyond the break radius tend to have redder, and probably older, stellar populations out there as well. This implies that star formation in these systems has retreated, at least for the present, to the inner regions of the galaxy (also seen by Zhang et al.\,2012a and referred to as ``outside-in'' disk growth) or inner stars have scattered to the outer parts over time, making the outer parts redder without changing the star formation pattern \citep[see also][]{pan15}.

\begin{deluxetable}{ccccccccc}
\tabletypesize{\scriptsize}
\tablecaption{Histogram of stellar mass fraction beyond \\the $V$-band break by color profile flavor
\label{tab-mass}}
\tablewidth{0pt}
\tablehead{\colhead{Mass} & \multicolumn{8}{c}{Color Profile Type} \\
\colhead{Fraction}  & \colhead{IIB} & \colhead{IIBF} & \colhead{IIBR} & \colhead{IIF} & \colhead{IIFR} & \colhead{IIR} & \colhead{III-B} & \colhead{III-I} 
}
\startdata
0--0.1    &  2  &  0  &  0  &   0  &   0  &  0  &   2  &   0 \\
0.1--0.2 &  4  &  2  &  1  &   1  &   0  &  1  &   2  &   0 \\
0.2--0.3 &  3  &  1  &  3  &   3  &   0  &  1  &   3  &   3 \\
0.3--0.4 &  5  &  1  &  0  &   3  &   4  &  3  &   2  &   1 \\
0.4--0.5 &  1  &  0  &  1  &   4  &   2  &  2  &   2  &   0 \\
0.5--0.6 &  0  &  0  &  1  &   2  &   1  &  3  &   0  &   1 \\
0.6--0.7 &  0  &  0  &  1  &   2  &   3  &  1  &   0  &   0 \\
0.7--0.8 &  0  &  0  &  0  &   2  &   2  &  1  &   0  &   0 \\
0.8--0.9 &  0  &  0  &  0  &   0  &   1  &  0  &   0  &   1 \\
0.9--1.0 &  0  &  0  &  0  &   0  &   0  &  0  &   0  &   0
\enddata
\end{deluxetable}

\section{DISCUSSION}
\subsection{Color Comparison to Spirals (B-V \& g'-r')}
As with the large spiral sample of BTP08, we find characteristic radial color trends for dwarfs of different surface brightness profile types where any breaks in the color trends occur near the breaks in the surface brightness profiles.  However, there is a larger number of color trends for dwarfs than for spirals since Type~II surface brightness profiles have significantly more than one radial color trend.

The color trends in dwarfs do not completely parallel those in spirals. The inner regions differ since all spirals have an initial bluing trend, explained by a radial decrease in age and metallicity (de Jong 1996), whereas only some Type~II dwarfs share this trend.  However, the outer trend in Type~IIIs is very similar in both spirals and dwarfs in all three colors explored here: a red trend between 0.5~$R_{br}$ and $R_{br}$ and then roughly constant beyond.  The spiral outer trend in Type~Is (i.e., constant) is seen in some dwarf Type~Is whereas they mainly get uniformly slightly redder with radius.  Note that the spiral study probes to $\sim$4.5 scale lengths whereas our dwarf average only probes to 3 scale lengths.  A significant fraction of Type~II dwarfs share the IIBR (blue then red) spiral ``U'' shape along with a single IIB (continuously blue) or IIR (continuously red) whereas smaller fractions are purely constant (IIF), flat then red (IIFR) or blue then flat (IIBF).  Note that the subset of Type~II profiles that have flat or increasing surface brightness with increasing radius (FI) have BR, F, FR, and R color trends but no B or BF color trends.  Recall that fainter ($-9 > M_B > -14$) Type~II dwarfs tend to have red (IIR) or ``U'' (IIBR) shaped colors and steeper color slopes than brighter ($-14 > M_B > -19$) Type~II dwarfs, which also have colors that become bluer (IIB) or stay the same (IIF) with increasing radius.

\subsection{Average Dwarf Profiles (and Bright vs.\,Faint)}
Recall that we did not average Type~II profiles with different color flavors earlier because we wanted to look for any possible differences between the different flavors.  BTP08 averaged virtually all their Type I spirals, Type II spirals, and Type III spirals, respectively, because virtually every individual galaxy of each subsample showed the characteristic color profile for its Type.  This was {\it not} the situation for our dwarf Type~II profiles.

However, there may be an important transition in dwarfs at $M_B \simeq$ -14.  For example, \citet{l+07} found a wider spread in H$\alpha$ equivalent widths in dwarfs fainter than -15 and \citet{z+12b} found that dwarfs fainter than -14.5 tend to have more three-dimensional than two-dimensional turbulence structures.  The distribution of our Type~II color trends further supports some change near this value.  Consequently, we split all the Type~II dwarfs used in Figures 1 and 2 (the broken-by-default, BBD, and single-by-default, SBD, analyses, respectively) into the ``bright'' and ``faint'' bins as with Figure~\ref{percent} and averaged all the bright Type~II profiles ($\mu_V$ and their corresponding $B-V$, $\log_{10} \Sigma$, $U-B$, and FUV$-$NUV profiles) after scaling the radii by the break location (for Type~II and III dwarfs) or one scale length (for Type~I dwarfs).  All profiles that were too short after being scaled radially as well as any originally classified as outliers were still not used.

\begin{figure*}
\epsscale{1.1}
\plotone{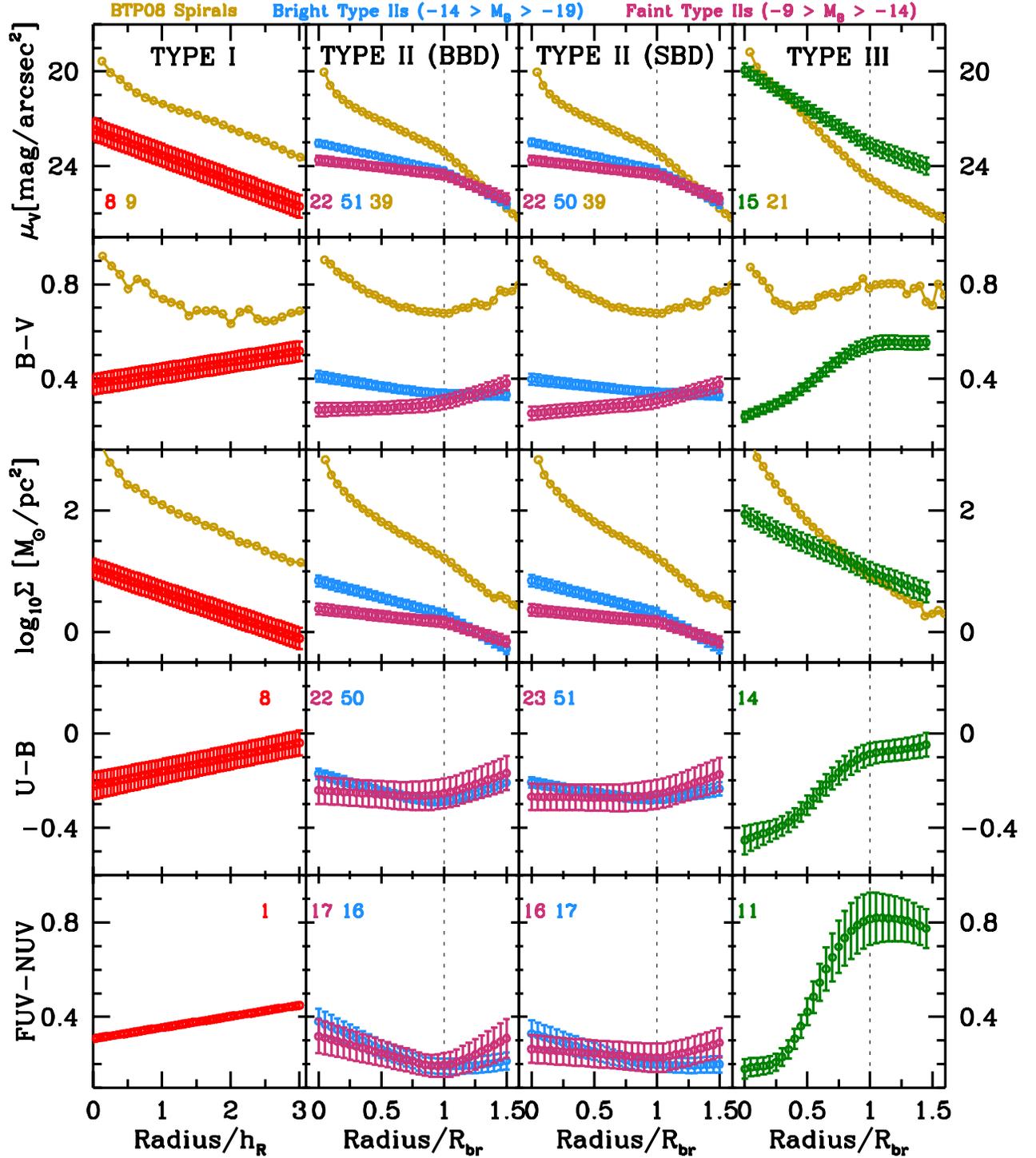}
\caption{Scaled and averaged profiles, but Type~II dwarfs have been split into two $M_B$ bins instead of being separated by flavor. BBD = broken-by-default and SBD = single-by-default.  {\it Top row}: $V$-band radial surface brightness profiles; {\it $2^{nd}$ row}: $B-V$ radial color profiles; {\it $3^{rd}$ row}: Surface mass density profiles obtained from applying a color-to-$M/L$ relationship (Herrmann et al., in preparation); {\it $4^{th}$ row}: $U-B$ radial color profiles; {\it Bottom row}: FUV$-$NUV radial color profiles.  The error bars show the uncertainty in the mean, i.e., $\sigma/\sqrt{N}$ where $\sigma$ is the standard deviation and $N$ (the colored numbers provided) is the number of galaxies used to calculate the average.  The gold data points show spiral results digitized from BTP08.  The red Type~I data are unchanged from Figures~1 and 2 and the dark green Type~III data are the averages of the SI and SB data from Figures~1 and 2. \label{NewAverages}}
\end{figure*}

Figure~\ref{NewAverages} presents the results.  The leftmost column of the Type~I profiles is exactly the same as in Figures~1 and 2, except all five rows have slightly smaller $y$ ranges and only the averages are being shown with the uncertainty in the mean on each radial data point.  The rightmost column is just the average of the Type~III dIms and BCDs with the stretched ``S'' presented in Figures~1 and 2.  The second column from the left shows the results for the bright Type~IIs (light blue) and the faint Type~IIs (red violet) for the BBD (broken-by-default) analysis whereas the third column from the left shows similar results, but for the SBD (single-by-default) analysis.  Happily, the BBD and SBD results are very similar, so the different fitting of the ambiguous color trends does not seem to be significant.  There are slight changes; for example, in $B-V$ the SBD curves are smoother (closer to a single line) whereas the BBD curves are closer to two lines, but that is not surprising.  The FUV$-$NUV plots are the most different, possibly due to the smaller sample sizes.

Since faint ($-9 > M_B > -14$) Type~II dwarfs tend to have IIR (continuous red) or IIBR (blue then red) color trends, it makes sense that when they are all averaged together the $B-V$ and $U-B$ results are IIFR (flattish then red) since the inner (outward from center) red and blue trends average to flattish and the outer (beyond the break) red and red trends average to red.  The dominance of IIBR trends in FUV$-$NUV causes the faint average to retain a IIBR shape.  The bright ($-14 > M_B > -19$) Type~II dwarfs additionally have IIB (continuous blue) and IIF (continuous flat) trends, so their averages in $B-V$ and FUV$-$NUV are roughly IIBF (blue then flat) but IIBR in $U-B$.

Since the Type~II dwarf $B-V$ averages are not IIBR, we see again that a kink generally {\it remains} in the surface mass density ($\Sigma$) profile for Type~II dwarfs but {\it disappears} in $\Sigma$ for Type~III dwarfs as in Section 5.1.  This is in {\it direct contrast} with the $\Sigma$ kink {\it disappearing} in Type~II spirals but {\it remaining} in Type~III spirals.  Moreover, the middle panels of the third row of Figure~\ref{NewAverages} show that the kink is strong in faint dwarfs, intermediate in bright dwarfs, and weak in spirals, indicating that surface mass density profiles may straighten with increasing galaxy mass.

\subsection{Comparison between Dwarf Groups}
{\it Sms:}  Almost all of the Sm galaxies have some (at least inner) {\it blue} trend (B, BR, or BF) in all three of the colors explored, when available.  The only Type~II exception is DDO~135 which has an ambiguous $U-B$ trend: either IIR (red) or a rare IIRB (red then blue) color profile.  The only Type~III Sm galaxy in our sample, UGC~11820, does not have the characteristic Type~III horizontally stretched ``S'' shape color trend; instead it has a flat (though generally bluish) trend in $B-V$ and a blue trend in $U-B$, which is more characteristic of Type~II Sm galaxies.  Note that all the Sm galaxies have $M_B$ brighter than -14.5.

{\it BCDs:}  In contrast to Sms, almost all BCDs have some (at least partial) {\it red} trend (R, BR, or SB) in all three of the colors explored, when available.  Since the BCD color sample is larger than the Sm sample (24 vs.\,17), it is not surprising that there are a few more exceptions: in $B-V$, Haro~8 is IFB (flat then blue), Mrk~600 is IIIFB, and Haro~29 is IIIBB (blue then blue) whereas in $U-B$, UGCA~290 is IIB (blue).  Additionally, Haro~36 is IIB in $B-V$ and IIF (flat) in $U-B$.  While the latter three color classifications are all typical of Type~II dwarfs, the former three are uncommon.  Moreover, all six of the BCDs with multiple profile types (for different passbands) have at least one Type~III surface brightness profile and all five with color information have at least one SB trend.

\subsection{Comparison between Colors}
Of the total 141 dwarfs in this analysis, 7 have no color information, 3 have only $B-V$ colors, 2 have only FUV$-$NUV colors, 73 have both $B-V$ and $U-B$ and 56 have all three colors.  Of the 56, 30 (54\%) dwarfs have the same color trend in all three colors (11 within ambiguous classifications), 20 (36\%) dwarfs have the same color trends in two colors (8 within ambiguous classification), and the remaining 6 (11\%) have completely different color trends between the three colors.  Of the 73 dwarfs with both $B-V$ and $U-B$, 43 (59\%) have the same color trend (16 within ambiguous classifications) whereas 30 (41\%) have different color trends.  There appears to be no dependence on color trend when the color trends are the same vs.\,when they differ.

\subsection{Outliers}
Recall that out of the 11 possible color trends (F, R, B, FR, FB, RF, RR, RB, BF, BR, BB), four were very uncommon: FB, RR, RB, and BB.

{\it Type~I Outliers:}  Although the ``normal'' trends for Type~I dwarfs were found to be IR (red) or IF (flat), a few Type~I dwarfs had other color trends.  DDO~34 is IB (blue) in $B-V$ and $U-B$ whereas Haro~8 is IFB (flat then blue) in $B-V$.  DDO~47 has the stretched ``S'' color trend characteristic of Type~III dwarfs in all three colors though it is Type~III in the ultraviolet but Type~I in the visible.  Lastly, though HS~0822+3542 has the typical IR (red) color trend, its $U-B$ colors are so drastically more blue than usual, so it was also tagged as an outlier and not included in the averaging analysis.

{\it Type~II Outliers:}  Between the numerous common Type~II color trends and ambiguous classifications, only three Type~II dwarfs are left as outliers.  DDO~69 is IIRR (red then red) in $B-V$, DDO~220 is IIRR in $U-B$, and DDO~216 has the stretched ``S'' color trend characteristic of Type~III dwarfs in both $B-V$ and $U-B$ even though it is a Type~II dwarf and has a IIBR (blue then red) profile in FUV$-$NUV.  Oddly enough all three Type~II outliers have some odd color trend with a {\it red} component.  Several profiles with ambiguous classifications were additionally tagged as outliers, but in only the broken-by-default classifications.  Since the single-by-default classification was ``normal,'' these dwarfs are not discussed here.

{\it Type~III Outliers:}  While the ``normal'' Type~III trend is a horizontally stretched ``S'' shape with some clear reddening trend and possibly some slight blue trend on either the inner or outer flattish section, five Type~III dwarfs have some stronger blue component.  DDO~120 is IIIRB/F in $U-B$, Haro~4 is IIIRB (red then blue) in $B-V$ and IIIRB/B in $U-B$, Haro~29 is IIIBB (blue then blue) in $B-V$, Mrk~600 is IIIFB (flat then blue) in $B-V$, and UGC~11820 (the only Type~III Sm) is IIIB (blue) in $U-B$.  Four more Type~III dwarfs instead have relatively constant colors: DDO~40 is IIIF (flat) in $B-V$ and $U-B$, DDO~68 is gently IIIR (red) in $B-V$, IIIF in $U-B$, and gently IIIB (blue) in FUV$-$NUV, NGC~3413 is gently IIIR in $B-V$ and $U-B$, and UGC~11820 (again the only Type~III Sm) is IIIF in $B-V$.

\subsection{Bars and/or Peculiar morphology}
Of the 129 galaxies with both $B-V$ and $U-B$ color data, 34 were classified as being barred and 23 had peculiar morphologies, both noted by \citet{he06}.  (Note that the barred and peculiar morphology subsamples have six overlapping dwarfs.)  Both subsamples contain a variety of the various color profile flavors, but both share the anomaly that a larger percentage tends to be IIB (blue) and fewer IIR (red) and IIBR (blue then red) with respect to the full sample.  There are very few IIBR color classifications in either subsample, except in the $U-B$ broken-by-default classifications.  Since 26/34 (76\%) of the barred sample and 21/23 (91\%) of the peculiar morphology sample have $M_B < -14$ and we noticed that brighter ($-14 > M_B > -19$) Type~II dwarfs tend to have more IIB or IIF (flat) color profiles than fainter ($-9 > M_B > -14$) Type~II dwarfs, the overabundance of brighter Type~II dwarfs in the barred and peculiar morphology subsamples may explain the overabundance of IIB classifications.

\section{CONCLUSIONS}
We have presented the color profiles for 141 dIm, BCD, and Sm galaxies with stellar surface brightness profiles presented in Paper~I.  The surface brightness profile types include Type~I (a single exponential profile), Type~II (an exponential with a break beyond which the profile becomes steeper), and Type~III (an exponential with a break beyond which the profile becomes shallower).  Here we fit the $B-V$, $U-B$, and FUV$-$NUV color profiles and classify the trends with radius according to whether they become blue, red, remain flat, or exhibit a combination of these trends. Thus, the Type~II surface brightness profiles are found to fit into the classes IIB, IIR, IIBR, IIF, IIFR, and IIBF.  Type~III profiles generally have horizontally stretched ``S''-shaped color profiles and are broken into III-B (also SB for ``S''-shaped BCD) and III-I (also SI, for ``S''-shaped dIm).  We use $B-V$ to convert $V$-band surface photometry to stellar surface mass density profiles, using the formalism for dwarf galaxies of Herrmann et al. (in preparation).  We find the following:

\begin{packed_enum}
\item In dwarf galaxies, the Type~II surface brightness breaks remain, although less pronounced, in the stellar surface mass density ($\Sigma$) profiles.  This is different from spirals where the $\Sigma$ breaks are significantly reduced and are interpreted as a change in stellar population. Thus, in dwarfs the break is, at least partly, a change in surface mass density.  The exception is the dwarf Type~II profiles that have ``U''-shaped color profiles (IIBR), like $all$ Type~II spirals, where the break mostly disappears in the $\Sigma$ profile of both $all$ Type~II spirals and dwarf IIBR profiles, but not $all$ Type~II dwarfs.  The straightness of $\Sigma$ in Type~II galaxies may depend on galaxy mass such that $\Sigma$ becomes straighter as galaxy mass increases from faint dwarfs to bright dwarfs to spirals.  Conversely, Type~III profiles are roughly continuous in $\Sigma$ profiles of {\it dwarfs} whereas a clear break remains in {\it spirals}.
\item Breaks in color profiles of dwarfs tend to occur roughly where the break occurs in the surface brightness profiles.
\item Dwarf galaxies with Type~I surface brightness profiles tend to become redder with radius (IR), whereas spirals tend to have a blueing trend that flattens beyond about 1.5 disk scale lengths.
\item Our sample of galaxies with Type~II surface brightness profiles exhibit six different flavors of color profiles, one of which is similar to the ``U'' shape (IIBR) that characterizes virtually all Type~II spiral color profiles.  The most common color trends for Type II dwarfs are those that get continuously bluer (IIB) or redder (IIR) with radius and those that start out getting bluer and then become redder with radius (IIBR).  Fainter ($-9 > M_B > -14$) dwarfs tend to get only redder with radius (IIR) or have ``U''-shaped colors (IIBR) while brighter ($-14 > M_B > -19$) dwarfs also commonly get bluer with radius (IIB) or remain flat (IIF). Type~II dwarf profiles that get bluer then stay constant (IIBF) and those that are flat before becoming redder farther out (IIFR) are less common.
\item For dwarf Type~II surface brightness profiles, the $B-V$ color at the break is roughly constant ($\sim$0.3).
\item Color profiles of Type~III surface brightness profiles are similar to those of spiral galaxies, exhibiting a horizontally stretched ``S'' shape in which the galaxy becomes progressively redder with a steep gradient in the middle to roughly the surface brightness break location and then is approximately constant in colors (III-I) or gradually varying (III-B).  The III-B flavor, seen in BCDs, has a stronger color gradient than the III-I flavor seen in dIm galaxies. Type~III galaxies tend to be bluer in color than Type~II galaxies in the inner parts and redder in the outer parts.
\item Sm-type dwarfs almost always have some blue radial color trend, and BCD-type dwarfs almost always have some reddening trend.  dIm galaxies cover the entire range of color profile flavors.
\item Type~II profiles that get bluer from the center outward in the inner galaxy have steeper inner surface brightness profiles (when the radii are scaled by $R_{br}$) than the Type~IIs that stay flat or get redder from the center outward in the inner disk.
\item The average stellar surface mass density at the break in the surface brightness profiles is about 1$-$2~$M_{\odot}$~pc$^{-2}$ for Type~II dwarfs but higher at 5.9~$M_{\odot}$~pc$^{-2}$ or 27~$M_{\odot}$~pc$^{-2}$ for Type~III BCDs and dIms, respectively.
\item Galaxies with Type~II profiles that become redder with radius tend to have more than 50\% of their stellar mass beyond the surface brightness break radius, while galaxies that have an inner blue trend (radially outward from the center) or Type~III profiles tend to have less than about 40\% of their stellar mass beyond the break radius.
\end{packed_enum}

What have we learned from the colors about the nature of the break in the stellar surface brightness profiles?  In most of the Type~II dwarfs the break is truly a change in surface mass density, but not necessarily a strong change.  The fact that changes in the color profiles tend to occur at or near the surface brightness breaks suggests that a change in stellar population, as occurs in spirals, may also enhance the surface brightness break.  In Paper I we found that the surface brightness breaks occur at about the same $V$-band surface brightness regardless of galaxy type or absolute magnitude. Here we see that the $B-V$ color at the break in Type~II dwarfs is also nearly constant at about $B-V\sim 0.3$. This reinforces the notion of a regularity in the nature of the change at the break radius among disk galaxies.

In Paper I of this series we summarized in Figure 18 the trends seen in the surface brightness profiles of different types and at different wavelengths. We suggested that the different profiles are revealing differences in evolutionary history:  Type~II galaxies are experiencing a star formation depletion in the inner disk so that the inner region profile becomes shallower; Type~III galaxies are experiencing an inner accretion that is increasing star formation over time there so that the inner region profile becomes steeper. The outer regions are not changing as much in either type but there is some evidence of Type~II profiles being consistent with outside-in shrinking of star formation in fainter dwarfs as found by Zhang et al.\ (2012a).

So how do the color profiles fit into this picture?  For Type~III galaxies the colors show an inner region that becomes gradually redder with radius, a middle region that becomes steeply redder with radius, and then an outer part that is relatively flat compared to the middle part of the curve.  Furthermore, Type~IIIs are bluer in the center and redder in the outer parts compared to Type~II galaxies.  These properties are consistent with the picture of star formation becoming centrally concentrated in Type~III systems.  Perhaps through accretion of gas into the center, Type~IIIs are currently, or have been recently, building their centers while the outer parts age.

Type~II galaxies, on the other hand, are more complex as a group.  In the picture presented in Paper I we would expect Type~II profiles to be systematically redder than Type~III in the inner parts and bluer in the outer parts, and this is observed.  On the other hand, if the central regions of Type~IIs are becoming depleted relative to the outer regions, we would expect a color profile that is flat or becomes bluer with radius. The brighter dIm galaxies with Type~II profiles do generally exhibit these kinds of color trends, but the fainter galaxies become steadily redder with radius or get bluer from the center out to the break and then redder beyond. In addition the redder trending Type~II profile is just as common as the bluer trending profile.  Yet, Type~II color profiles that are flat or red-trending in the inner galaxy are also flatter in surface brightness than those that are blue as one might expect in a central region that is winding down with time in star formation.  However, the Type~IIR galaxies typically have a significant stellar mass beyond the break radius with redder, and probably older, stellar populations out there. This implies that star formation in these systems has retreated, at least for the present, to the inner regions of the galaxy, contrary to the straw picture presented in Paper I. Perhaps the mixed nature of color profiles for Type~II systems are telling us that these galaxies are caught in a variety of evolutionary stages.

\acknowledgments
We thank the anonymous reviewer for useful suggestions and gratefully acknowledge partial funding for this research from the National Science Foundation grants AST-0707563 to DAH and AST-0707426 to BGE.  DAH also appreciates support from the Lowell Research Fund and is grateful to John and Meg Menke for a donation to Lowell Observatory that covered part of the page charges.  KAH also acknowledges funding from Pennsylvania State Mont Alto that enabled extended visits to Lowell Observatory where much of this work was carried out.


\begin{thebibliography}{}
\bibitem[Azzollini et~al.(2008)]{atb08} Azzollini, R., Trujillo, I., \& Beckman, J.\,E.\,2008, \aj, 684, 1026
\bibitem[Bakos et~al.(2008)]{btp08} Bakos, J., Trujillo, I., \& Pohlen, M.\,2008, \apj, 683, L103 (BTP08)
\bibitem[Bell et~al.(2003)]{b+03} Bell, E.\,F., McIntosh, D.\,H., Katz, N., \& Weinberg, M.\,D.\,2003, \apjs, 149, 289
\bibitem[de Jong(1996)]{dJ96} de Jong, R.\,S.\,1996, \aap, 313, 377
\bibitem[de Vaucouleurs et~al.(1991)]{RC3} de Vaucouleurs, G., de Vaucouleurs, A., Corwin, H., et~al.\,1991, Third Reference Catalogue of Bright Galaxies (New York: Springer) (RC3)
\bibitem[Erwin et~al.(2005)]{ebp05} Erwin, P., Beckman, J.\,E., \& Pohlen, M.\,2005, \apj, 626, L81
\bibitem[Erwin et~al.(2008)]{epb08} Erwin, P., Pohlen, M., \& Beckman, J.\,E.\,2008, \aj, 135, 20
\bibitem[Fazio et~al.(2004]{fazio04} Fazio, G.\,G., Hora, J.\,L., Allen, L.\,E., et al.\,2004, \apjs, 154, 10
\bibitem[Guti\'{e}rrez et~al.(2011)]{geab11} Guti\'{e}rrez, L., Erwin, P., Aladro, R., \& Beckman, J.\,E.\,2011, \aj, 142, 145
\bibitem[Herrmann et~al.(2013)]{PaperI} Herrmann, K.\,A., Hunter, D.\,A., \& Elmegreen, B.\,G.\, 2013, \aj, 146, 104 (Paper~I)
\bibitem[Hodge \& Hitchcock(1966)]{hh66} Hodge, P.\,W., \& Hitchcock, J.\,L.\,1966, \pasp, 78, 79
\bibitem[Hunter \& Elmegreen(2004)]{he04} Hunter, D.\,A., \& Elmegreen, B.\,G.\,2004, \apj, 128, 2170
\bibitem[Hunter \& Elmegreen(2006)]{he06} Hunter, D.\,A., \& Elmegreen, B.\,G.\,2006, \apjs, 162, 49
\bibitem[Hunter et~al.(2010)]{hel10} Hunter, D.\,A., Elmegreen, B.\,G., \& Ludka, B.\,C.\,2010, \aj, 139, 447
\bibitem[Hunter et~al.(2006)]{hem06} Hunter, D.\,A., Elmegreen, B.\,G., \& Martin, E.\,2006, \aj, 132, 801
\bibitem[Hunter et~al.(2011)]{deep} Hunter, D.\,A., Elmegreen, B.\,G., Oh, S.\-H., et al.\,2011, \aj, 142, 121
\bibitem[Hunter et~al.(2012)]{LTdata} Hunter, D.\,A., Ficut-Vicas, D., Ashley, T., et al.\,2012, \aj, 144, 134
\bibitem[Kroupa(2001)]{k01} Kroupa, P.\,2001, \mnras, 322, 231
\bibitem[Lee et~al.(2007)]{l+07} Lee, J.\,C., Kennicutt, R.\,C., Funes, S.\,J., et al.\,2007, \apj, 671, L113
\bibitem[Leitherer \& Chen(2009)]{SB99c} Leitherer, C., \& Chen, J.\,2009, New Astronomy, 14, 356
\bibitem[Leitherer et~al.(2014)]{SB99d} Leitherer, C., Ekstr\"{o}m, S., Meynet, G., et al.\,2014, \apjs, 212, 14
\bibitem[Leitherer et~al.(1999)]{SB99a} Leitherer, C., Schaerer, D., Goldader, J.\,D., et al.\,1999, \apjs, 123, 3
\bibitem[Martin et~al.(2005)]{martin05} Martin, D.\,C., Fansom, J., Schiminovich, D., et al.\,2005, \apj, 619, L1
\bibitem[Mart\'{i}nez-Serrano et~al.(2009)]{ms+09} Mart\'{i}nez-Serrano, F.\,J., Serna, A., Dom\'{e}nech-Moral, M., \& Dom\'{i}nguez-Tenreiro, R.\,2009, \apj, 705, L133
\bibitem[Pan et al.(2015)]{pan15} Pan, Z., Li, J., Lin, W., et al.\,2015, \apjl, 804, L42
\bibitem[P\'{e}rez(2004)]{p04} P\'{e}rez, I.\,2004, \aap, 427, L17
\bibitem[Pohlen \& Trujillo(2006)]{pt06} Pohlen, M., \& Trujillo, I.\,2006, \aap, 454, 759
\bibitem[Press et~al.(2002)]{NumRec} Press, W.\,H., Teukolsky, S.\,A., Vetterling, W.\,T., \& Flannery, B.\,P.\,2002, Numerical Recipes in C++: The Art of Scientific Computing, Second Edition, (New York, NY: Cambridge University Press)
\bibitem[Ro\u{s}kar et~al.(2008a)]{r+08a} Ro\u{s}kar, R., Debattista, V.\,P., Quinn, T.\,R., Stinson, G.\,S., \& Wadsley, J.\, 2008b, \apj, 684, L79
\bibitem[Ro\u{s}kar et~al.(2008b)]{r+08b} Ro\u{s}kar, R., Debattista, V.\,P., Stinson, G.\,S., et~al.\, 2008b, \apj, 675, L65
\bibitem[S\'{a}nchez-Bl\'{a}zquez et~al.(2009)]{sb+09} S\'{a}nchez-Bl\'{a}zquez, P., Courty, S., Gibson, B.\,K., \& Brook, C.\,B.\,2009, \mnras, 398, 591
\bibitem[Smith et~al.(2002)]{SDSSfilters} Smith, J.\,A., Tucker, D.\,L., Kent, S., et~al.\,, 2002, \aj, 123, 2121
\bibitem[Trujillo et~al.(2009)]{t+09} Trujillo, I., Azzollini, R., Bakos, J., Beckman, J.\,E., \& Pohlen, M.\,2009, in IAU Symp.\,254, The Galaxy Disk in Cosmological Context, ed.\,J.\,Andersen, J.\,Bland-Hawthorn, \& B.\,Nordstr\"{o}m (Cambridge: Cambridge Univ.\,Press), 127
\bibitem[van den Bergh(1988)]{vdB88} van den Bergh, S.\,1988, \pasp, 100, 344
\bibitem[Vazquez \& Leitherer(2005)]{SB99b} V\'{a}zquez, G.\,A., \& Leitherer, C.\,2005, \apj, 621, 695
\bibitem[Walter et~al.(2008)]{THINGS} Walter, F., Brinks, E., de Blok, W.\,J.\,G., et~al.\,2008, \aj, 136, 2563
\bibitem[Zhang et~al.(2012a)]{z+12a} Zhang, H.-X., Hunter, D.\,A., Elmegreen, B.\,G., Gao, Y., \& Schruba, A.\,2012a, \aj, 143, 47
\bibitem[Zhang et~al.(2012b)]{z+12b} Zhang, H.-X., Hunter,  D.\,A., \&  Elmegreen, B.\,G.\,2012b, \apj, 754, 29
\end{thebibliography}
\end{document}